\documentclass[twocolumn,twocolappendix]{aastex63}
\hypersetup{linkcolor=blue,citecolor=blue,filecolor=blue,urlcolor=blue}
\usepackage{amsmath}
\newcommand{\gaia}{\emph{Gaia}}

\newcommand{\tess}[1]{\emph{TESS}#1}

\newcommand{\teff}[1]{$T_{\text{eff}}$#1}
\newcommand{\prot}[1]{$P_{\text{rot}}$#1}
\newcommand{\logg}[1]{$\log{g}$#1}
\newcommand{\vsini}[1]{$v\sin{i}$#1}
\newcommand{\mps}[1]{m s$^{-1}$#1}
\newcommand{\teq}[1]{$T_{\text{eq}}$#1}
\newcommand{\Rearth}[1]{R$_{\oplus}$#1}
\newcommand{\Rsun}[1]{R$_{\odot}$#1}
\newcommand{\Mearth}[1]{M$_{\oplus}$#1}
\newcommand{\Msun}[1]{M$_{\odot}$#1}
\newcommand{\name}[1]{TOI-1235#1}
\defcitealias{cloutier20}{CM19}
\defcitealias{martinez19}{M19}

\shortauthors{Cloutier et al.}
\shorttitle{\tess{} discovery of \name{} b}

\begin{document}
\title{\name{} b: a keystone super-Earth for testing radius valley emergence models around early M dwarfs}

\suppressAffiliations
\author[0000-0001-5383-9393]{Ryan Cloutier}
\affiliation{Center for Astrophysics $|$ Harvard \& Smithsonian, 60 Garden
  Street, Cambridge, MA, 02138, USA}

\author[0000-0001-8812-0565]{Joseph E. Rodriguez}  
\affiliation{Center for Astrophysics $|$ Harvard \& Smithsonian, 60 Garden
  Street, Cambridge, MA, 02138, USA}

\author{Jonathan Irwin}  
\affiliation{Center for Astrophysics $|$ Harvard \& Smithsonian, 60 Garden
  Street, Cambridge, MA, 02138, USA}

\author[0000-0002-9003-484X]{David Charbonneau}  
\affiliation{Center for Astrophysics $|$ Harvard \& Smithsonian, 60 Garden
  Street, Cambridge, MA, 02138, USA}

\author[0000-0002-3481-9052]{Keivan G. Stassun}  
\affiliation{Department of Physics \& Astronomy, Vanderbilt University, 
6301 Stevenson Center Lane, Nashville, TN, 37235, USA}

\author[0000-0001-7254-4363]{Annelies Mortier}  
\affiliation{Astrophysics Group, Cavendish Laboratory, University of Cambridge,
  J.J. Thomson Avenue, Cambridge CB3 0HE, UK}

\author{David W. Latham}  
\affiliation{Center for Astrophysics $|$ Harvard \& Smithsonian, 60 Garden
  Street, Cambridge, MA, 02138, USA}

\author[0000-0002-0531-1073]{Howard Isaacson} 
\affiliation{501 Campbell Hall, University of California at Berkeley, Berkeley, CA 94720, USA}
\affiliation{Centre for Astrophysics, University of Southern Queensland,
Toowoomba, QLD, Australia}

\author[0000-0001-8638-0320]{Andrew W. Howard}  
\affiliation{Department of Astronomy, California Institute of Technology, Pasadena, CA 91125, USA}

\author{St\'ephane Udry}  
\affiliation{Observatoire Astronomique de l'Universit\'e de Gen\`eve, 51 chemin
  des Maillettes, 1290 Versoix, Switzerland}

\author[0000-0001-8749-1962]{Thomas G. Wilson}  
\affiliation{School of Physics and Astronomy, University of St Andrews, North
Haugh, St Andrews, Fife, KY16 9SS, UK}

\author{Christopher A. Watson}  
\affiliation{Astrophysics Research Centre, School of Mathematics and Physics,
Queen's University Belfast, Belfast, BT7 1NN, UK}

\author[0000-0002-4445-1845]{Matteo Pinamonti} 
\affiliation{INAF - Osservatorio Astrofisico di Torino, Strada Osservatorio 20,
  Pino Torinese (To) 10025, Italy}

\author[0000-0003-4047-0771]{Florian Lienhard}  
\affiliation{Astrophysics Group, Cavendish Laboratory, University of Cambridge,
  J.J. Thomson Avenue, Cambridge CB3 0HE, UK}

\author{Paolo Giacobbe} 
\affiliation{INAF - Osservatorio Astrofisico di Torino, Strada Osservatorio 20,
  Pino Torinese (To) 10025, Italy}

\author[0000-0002-4308-2339]{Pere Guerra}  
\affiliation{Observatori Astron\`omic Albany\`a, Cam\'i de Bassegoda S/N, 
  Albany\`a 17733, Girona, Spain}

\author[0000-0001-6588-9574]{Karen A. Collins}  
\affiliation{Center for Astrophysics $|$ Harvard \& Smithsonian, 60 Garden
  Street, Cambridge, MA, 02138, USA}

\author[0000-0001-6637-5401]{Allyson Beiryla}  
\affiliation{Center for Astrophysics $|$ Harvard \& Smithsonian, 60 Garden
  Street, Cambridge, MA, 02138, USA}

\author[0000-0002-9789-5474]{Gilbert A. Esquerdo}  
\affiliation{Center for Astrophysics $|$ Harvard \& Smithsonian, 60 Garden
  Street, Cambridge, MA, 02138, USA}


\author[0000-0003-0593-1560]{Elisabeth Matthews}   
\affiliation{Department of Earth, Atmospheric and Planetary Sciences, and
Kavli Institute for Astrophysics and Space Research, Massachusetts Institute
of Technology, Cambridge, MA 02139, USA}

\author[0000-0001-7233-7508]{Rachel A. Matson}   
\affiliation{U.S. Naval Observatory, Washington, DC 20392, USA}

\author[0000-0002-2532-2853]{Steve B. Howell}  
\affiliation{NASA Ames Research Center, Moffett Field, CA, 94035, USA}

\author{Elise Furlan}  
\affiliation{Caltech/IPAC, 1200 E. California Blvd. Pasadena, CA 91125, USA}

\author{Ian J. M. Crossfield}  
\affiliation{Department of Physics \& Astronomy, University of Kansas, 1082
  Malott, 1251 Wescoe Hall Dr. Lawrence, KS, 66045, USA}

\author[0000-0001-6031-9513]{Jennifer G. Winters}  
\affiliation{Center for Astrophysics $|$ Harvard \& Smithsonian, 60 Garden
  Street, Cambridge, MA, 02138, USA}

\author[0000-0001-8838-3883]{Chantanelle Nava}  
\affiliation{Center for Astrophysics $|$ Harvard \& Smithsonian, 60 Garden
  Street, Cambridge, MA, 02138, USA}

\author{Kristo Ment}  
\affiliation{Center for Astrophysics $|$ Harvard \& Smithsonian, 60 Garden
  Street, Cambridge, MA, 02138, USA}

\author{Eric D. Lopez}  
\affiliation{NASA Goddard Space Flight Center, 8800 Greenbelt Rd, Greenbelt,
  MD 20771, USA}
\affiliation{GSFC Sellers Exoplanet Environments Collaboration, NASA Goddard
Space Flight Center, Greenbelt, MD 20771}

\author{George Ricker}
\affiliation{Department of Earth, Atmospheric and Planetary Sciences, and 
Kavli Institute for Astrophysics and Space Research, Massachusetts Institute 
of Technology, Cambridge, MA 02139, USA}

\author[0000-0001-6763-6562]{Roland Vanderspek}
\affiliation{Department of Earth, Atmospheric and Planetary Sciences, and 
Kavli Institute for Astrophysics and Space Research, Massachusetts Institute 
of Technology, Cambridge, MA 02139, USA}

\author[0000-0002-6892-6948]{Sara Seager}
\affiliation{Department of Physics and Kavli Institute for Astrophysics and
Space Research, Massachusetts Institute of Technology, Cambridge, MA 02139, USA}
\affiliation{Department of Earth, Atmospheric and Planetary Sciences,
Massachusetts Institute of Technology, Cambridge, MA 02139, USA}
\affiliation{Department of Aeronautics and Astronautics, MIT, 77 Massachusetts
Avenue, Cambridge, MA 02139, USA}


\author[0000-0002-4715-9460]{Jon M. Jenkins}
\affiliation{NASA Ames Research Center, Moffett Field, CA, 94035, USA}

\author[0000-0002-8219-9505]{Eric B. Ting}  
\affiliation{NASA Ames Research Center, Moffett Field, CA, 94035, USA}

\author[0000-0002-1949-4720]{Peter Tenenbaum}  
\affiliation{NASA Ames Research Center, Moffett Field, CA, 94035, USA}

\author[0000-0002-7504-365X]{Alessandro Sozzetti}  
\affiliation{INAF - Osservatorio Astrofisico di Torino, Strada Osservatorio 20,
  Pino Torinese (To) 10025, Italy}

\author[0000-0001-5401-8079]{Lizhou Sha}  
\affiliation{Department of Physics and Kavli Institute for Astrophysics and
Space Research, Massachusetts Institute of Technology, Cambridge, MA 02139, USA}

\author{Damien S\'egransan}  
\affiliation{Observatoire Astronomique de l'Universit\'e de Gen\`eve, 51 chemin
  des Maillettes, 1290 Versoix, Switzerland}

\author[0000-0001-5347-7062]{Joshua E. Schlieder}  
\affiliation{NASA Goddard Space Flight Center, 8800 Greenbelt Rd, Greenbelt,
  MD 20771, USA}

\author[0000-0001-7014-1771]{Dimitar Sasselov}  
\affiliation{Center for Astrophysics $|$ Harvard \& Smithsonian, 60 Garden
  Street, Cambridge, MA, 02138, USA}

\author[0000-0001-8127-5775]{Arpita Roy}  
\affiliation{Department of Astronomy, California Institute of Technology, Pasadena, CA 91125, USA}

\author[0000-0003-0149-9678]{Paul Robertson}  
\affiliation{Department of Physics \& Astronomy, University of California Irvine, Irvine, CA 92697, USA}

\author{Ken Rice}  
\affiliation{SUPA, Institute for Astronomy, University of Edinburgh, Blackford
  Hill, Edinburgh, EH9 3HJ, Scotland, UK}

\author[0000-0003-1200-0473]{Ennio Poretti}  
\affiliation{Fundaci\'on Galileo Galilei-INAF, Rambla Jos\'e Ana Fernandez
P\'erez 7, 38712 Bre\~{n}a Baja, TF, Spain}
\affiliation{INAF-Osservatorio Astronomico di Brera, via E. Bianchi 46, 23807
Merate (LC), Italy}

\author{Giampaolo Piotto}  
\affiliation{Dip. di Fisica e Astronomia Galileo Galilei - Universit\`a di
Padova, Vicolo dell'Osservatorio 2, 35122, Padova, Italy}

\author{David Phillips}  
\affiliation{Center for Astrophysics $|$ Harvard \& Smithsonian, 60 Garden
  Street, Cambridge, MA, 02138, USA}

\author[0000-0002-3827-8417]{Joshua Pepper}  
\affiliation{Lehigh University, Department of Physics, 16 Memorial Drive East,
Bethlehem, PA, 18015, USA}

\author{Francesco Pepe}  
\affiliation{Observatoire Astronomique de l'Universit\'e de Gen\`eve, 51 chemin
  des Maillettes, 1290 Versoix, Switzerland}

\author[0000-0002-1742-7735]{Emilio Molinari}  
\affiliation{INAF - Osservatorio Astronomico di Cagliari, via della Scienza 5,
09047, Selargius, Italy}

\author[0000-0003-4603-556X]{Teo Mocnik}  
\affiliation{Gemini Observatory Northern Operations, 670 N. A'ohoku Place, Hilo, HI 96720, USA}

\author[0000-0002-9900-4751]{Giuseppina Micela}  
\affiliation{INAF - Osservatorio Astronomico di Palermo, Piazza del Parlamento
1, I-90134 Palermo, Italy}


\author{Michel Mayor}  
\affiliation{Observatoire Astronomique de l'Universit\'e de Gen\`eve, 51 chemin
  des Maillettes, 1290 Versoix, Switzerland}

\author{Aldo F. Martinez Fiorenzano}  
\affiliation{Fundaci\'on Galileo Galilei-INAF, Rambla Jos\'e Ana Fernandez
P\'erez 7, 38712 Bre\~{n}a Baja, TF, Spain}

\author{Franco Mallia}  
\affiliation{Campo Catino Astronomical Observatory, Regione Lazio, Guarcino
(FR), 03010 Italy}

\author[0000-0001-8342-7736]{Jack Lubin} 
\affiliation{Department of Physics \& Astronomy, University of California Irvine, Irvine, CA 92697, USA}

\author{Christophe Lovis}  
\affiliation{Observatoire Astronomique de l'Universit\'e de Gen\`eve, 51 chemin
  des Maillettes, 1290 Versoix, Switzerland}

\author[0000-0003-3204-8183]{Mercedes L\'opez-Morales}  
\affiliation{Center for Astrophysics $|$ Harvard \& Smithsonian, 60 Garden
  Street, Cambridge, MA, 02138, USA}

\author{Molly R. Kosiarek}  
\affiliation{Department of Astronomy and Astrophysics, University of California, Santa Cruz, CA 95064, USA}

\author[0000-0003-0497-2651]{John F. Kielkopf}  
\affiliation{Department of Physics and Astronomy, University of Louisville,
Louisville, KY 40292, USA}

\author[0000-0002-7084-0529]{Stephen R. Kane}  
\affiliation{Department of Earth and Planetary Sciences, University of California, Riverside, CA 92521, USA}

\author[0000-0002-4625-7333]{Eric L. N. Jensen}  
\affiliation{Dept. of Physics \& Astronomy, Swarthmore College, Swarthmore PA
  19081, USA}

\author{Giovanni Isopi}  
\affiliation{Campo Catino Astronomical Observatory, Regione Lazio, Guarcino
(FR), 03010 Italy}

\author[0000-0001-8832-4488]{Daniel Huber}  
\affiliation{Institute for Astronomy, University of Hawai`i, 2680 Woodlawn Drive, Honolulu, HI 96822, USA}

\author[0000-0002-0139-4756]{Michelle L. Hill} 
\affiliation{Department of Earth and Planetary Sciences, University of California, Riverside, CA 92521, USA}

\author{Avet Harutyunyan} 
\affiliation{Fundaci\'on Galileo Galilei-INAF, Rambla Jos\'e Ana Fernandez
P\'erez 7, 38712 Bre\~{n}a Baja, TF, Spain}

\author{Erica Gonzales}  
\affiliation{Department of Astronomy and Astrophysics, University of California, Santa Cruz, CA 95064, USA}

\author[0000-0002-8965-3969]{Steven Giacalone}  
\affiliation{501 Campbell Hall, University of California at Berkeley, Berkeley, CA 94720, USA}

\author[0000-0003-4702-5152]{Adriano Ghedina}  
\affiliation{Fundaci\'on Galileo Galilei-INAF, Rambla Jos\'e Ana Fernandez
P\'erez 7, 38712 Bre\~{n}a Baja, TF, Spain}

\author{Andrea Ercolino}  
\affiliation{Campo Catino Astronomical Observatory, Regione Lazio, Guarcino
(FR), 03010 Italy}

\author{Xavier Dumusque}  
\affiliation{Observatoire Astronomique de l'Universit\'e de Gen\`eve, 51 chemin
  des Maillettes, 1290 Versoix, Switzerland}

\author[0000-0001-8189-0233]{Courtney D. Dressing}
\affiliation{501 Campbell Hall, University of California at Berkeley, Berkeley, CA 94720, USA}

\author{Mario Damasso}  
\affiliation{INAF - Osservatorio Astrofisico di Torino, Strada Osservatorio 20,
  Pino Torinese (To) 10025, Italy}

\author[0000-0002-4297-5506]{Paul A. Dalba}
\altaffiliation{NSF Astronomy and Astrophysics Postdoctoral Fellow}
\affiliation{Department of Earth and Planetary Sciences, University of California, Riverside, CA 92521, USA}

\author[0000-0003-1784-1431]{Rosario Cosentino}  
\affiliation{Fundaci\'on Galileo Galilei-INAF, Rambla Jos\'e Ana Fernandez
P\'erez 7, 38712 Bre\~{n}a Baja, TF, Spain}

\author[0000-0003-2239-0567]{Dennis M. Conti}  
\affiliation{American Association of Variable Star Observers, 49 Bay State
Road, Cambridge, MA 02138, USA}

\author[0000-0001-8020-7121]{Knicole D. Col\'{o}n}  
\affiliation{NASA Goddard Space Flight Center, Exoplanets and Stellar Astrophysics Laboratory (Code 667), Greenbelt, MD 20771, USA}  

\author[0000-0003-2781-3207]{Kevin I. Collins}  
\affiliation{George Mason University, 4400 University Drive, Fairfax, VA,
22030 USA}

\author{Andrew Collier Cameron}  
\affiliation{School of Physics and Astronomy, University of St Andrews, North
Haugh, St Andrews, Fife, KY16 9SS, UK}

\author{David Ciardi}  
\affiliation{Caltech/IPAC, 1200 E. California Blvd. Pasadena, CA 91125, USA}

\author{Jessie Christiansen} 
\affiliation{Caltech/IPAC, 1200 E. California Blvd. Pasadena, CA 91125, USA}

\author[0000-0003-1125-2564]{Ashley Chontos}  
\altaffiliation{NSF Graduate Research Fellow}
\affiliation{Institute for Astronomy, University of Hawai`i, 2680 Woodlawn Drive, Honolulu, HI 96822, USA}

\author{Massimo Cecconi}  
\affiliation{Fundaci\'on Galileo Galilei-INAF, Rambla Jos\'e Ana Fernandez
P\'erez 7, 38712 Bre\~{n}a Baja, TF, Spain}

\author[0000-0003-1963-9616]{Douglas A. Caldwell} 
\affiliation{NASA Ames Research Center, Moffett Field, CA, 94035, USA}

\author{Christopher Burke}  
\affiliation{Kavli Institute for Astrophysics and Space Research, Massachusetts Institute
of Technology, Cambridge, MA 02139, USA}

\author{Lars Buchhave}  
\affiliation{DTU Space, National Space Institute, Technical University of 
Denmark, Elektrovej 328, DK-2800 Kgs. Lyngby, Denmark}

\author{Charles Beichman}  
\affiliation{NASA Exoplanet Science Institute, Infrared Processing \& Analysis
  Center, Jet Propulsion Laboratory, California Institute of Technology,
  Pasadena CA, 91125, USA}

\author[0000-0003-0012-9093]{Aida Behmard}  
\altaffiliation{NSF Graduate Research Fellow}
\affiliation{Division of Geological and Planetary Science, California Institute of Technology, Pasadena, CA 91125, USA}

\author[0000-0001-7708-2364]{Corey Beard}  
\affiliation{Department of Physics \& Astronomy, University of California Irvine, Irvine, CA 92697, USA}

\author[0000-0001-8898-8284]{Joseph M. Akana Murphy}
\altaffiliation{NSF Graduate Research Fellow}
\affiliation{Department of Astronomy and Astrophysics, University of California, Santa Cruz, CA 95064, USA}

\correspondingauthor{Ryan Cloutier}
\email{ryan.cloutier@cfa.harvard.edu}

\begin{abstract}
  Small planets on close-in orbits tend to exhibit envelope mass fractions of
  either effectively zero or up to a few percent depending on their
  size and orbital period. Models of thermally-driven atmospheric mass loss
  and of terrestrial planet formation in a gas-poor environment make distinct
  predictions regarding the location of this rocky/non-rocky transition in
  period-radius space.
  Here we present the confirmation of \name{} b ($P=3.44$ days, $r_p=1.738^{+0.087}_{-0.076}$ \Rearth{)}, a planet whose size and
  period are intermediate between the competing model predictions thus making
  the system an important test case for emergence models of the rocky/non-rocky
  transition around early M dwarfs ($R_s=0.630\pm 0.015$ \Rsun{,}
  $M_s=0.640\pm 0.016$ \Msun{)}. We confirm the \tess{} planet discovery using
  reconnaissance spectroscopy,
  ground-based photometry, high-resolution imaging, and a set of 38
  precise radial-velocities from HARPS-N and HIRES.
  We measure a planet mass of $6.91^{+0.75}_{-0.85}$ \Mearth{,} which
  implies an iron core mass fraction of $20^{+15}_{-12}$\% in the absence
  of a gaseous envelope.
  The bulk composition of \name{} b is therefore consistent with being
  Earth-like and we
  constrain a H/He envelope mass fraction to be $<0.5$\% at 90\% confidence.
  Our results are consistent with model predictions
  from thermally-driven atmospheric mass loss but not with gas-poor formation,
  suggesting that the former class of processes remain efficient at
  sculpting close-in planets around early M dwarfs. Our RV analysis also reveals
  a strong periodicity close to the first harmonic of the
  photometrically-determined stellar rotation period that we treat as stellar
  activity, despite other lines of evidence favoring a planetary origin
  ($P=21.8^{+0.9}_{-0.8}$ days, $m_p\sin{i}=13.0^{+3.8}_{-5.3}$ \Mearth{)} that
  cannot be firmly ruled out by our data.
\end{abstract}

\section{Introduction}
The occurrence rate distribution of close-in planets features a dearth of
planets between 1.7-2.0 \Rearth{} around Sun-like stars
\citep[\teff{} $>4700$ K;][]{fulton17,fulton18,mayo18}
and between 1.4-1.7 \Rearth{} around mid-K to
mid-M dwarfs \citep[\teff{} $<4700$ K;][]{cloutier20}.
The so-called radius valley likely emerges due to the existence of
a transition from primarily rocky planets to larger non-rocky planets that host
extended H/He envelopes up to a few percent by mass
\citep{weiss14,rogers15,dressing15b}. Furthermore, the exact location of the
rocky/non-rocky transition around both Sun-like and lower mass stars is known
to be period-dependent
\citep{vaneylen18,martinez19,wu19,cloutier20}, with the model-predicted slope of
the period dependence varying between competing physical models that
describe potential pathways for the radius valley's emergence.
One class of models rely on thermal heating to
drive atmospheric escape. For example, photoevaporation, wherein a planet's
primordial atmosphere is stripped by XUV photons from the host star during the
first 100 Myrs \citep{owen13,jin14,lopez14,chen16,owen17,jin18,lopez18,wu19},
predicts that the slope of the radius valley should vary with orbital period
as $r_{p,\text{valley}}\propto P^{-0.15}$ \citep{lopez18}. A similar slope of
$r_{p,\text{valley}}\propto P^{-0.13}$ \citep{gupta20} is
predicted by internally-driven thermal atmospheric escape models
via the core-powered
mass loss mechanism \citep{ginzburg18,gupta19,gupta20}. However, 
if instead the radius valley emerges from the superposition of rocky and
non-rocky planet populations, wherein the former are formed at late times in a
gas-poor
environment \citep{lee14,lee16,lopez18}, then the period-dependence of the
radius valley should have the opposite sign: $r_{p,\text{valley}} \propto P^{0.11}$
\citep{lopez18}. These
distinct slope predictions naturally carve out a subspace in period-radius space
wherein knowledge of planetary bulk compositions can directly constrain the
applicability of each class of model
\citep[Fig. 15,][hereafter \citetalias{cloutier20}]{cloutier20}.
This is because within that subspace, and at $\lesssim 23.5$ days
\citepalias{cloutier20}, thermally-driven
mass loss models predict that planets will be rocky whereas the gas-poor
formation model predicts non-rocky planets.
Therefore, populating this subspace with planets
with known bulk compositions will inform the prevalence of each model as
a function of host stellar mass. 

Since the commencement of its prime mission in July 2018, NASA's Transiting
Exoplanet Survey Satellite \citep[\tess{;}][]{ricker15} has uncovered a number
of transiting planet candidates whose orbital periods and radii lie within the
aforementioned subspace. These planets are valuable targets to conduct tests of
competing radius valley emergence models across a range of stellar
masses through the characterization of their bulk compositions using precise
radial-velocity measurements. Here we present the confirmation of one such
planet from \tess{:} \name{} b (TIC 103633434.01). Our analysis
includes the mass measurement of \name{} b from 38 radial-velocity
observations from HARPS-N and HIRES. Our RV observations also reveal a second
signal at 22 days that is suggestive of arising from stellar rotation, although
some counter-evidence favors a planetary interpretation that cannot be firmly
ruled out by our data.

In Sect.~\ref{sect:star} we present the properties of the host star \name{}. In
Sect.~\ref{sect:observations} we present the \tess{} light curve and our
suite of follow-up observations including a measurement of the stellar
rotation period from archival photometric monitoring.
In Sect.~\ref{sect:analysis} we present our data analysis and results.
We conclude with a discussion and a summary of our results in
Sects.~\ref{sect:discussion} and~\ref{sect:summary}.

\section{Stellar Characterization} \label{sect:star}
\name{} (TIC 103633434, TYC 4384-1735-1, Gaia DR2 1070387905514406400)
is an early M dwarf located in the northern sky at a distance of
$39.635\pm 0.047$
pc\footnote{The \gaia{} DR2 parallax is corrected by $+0.08$ mas to account
  for the systematic offset reported by \cite{stassun18a}.}
\citep{gaia18,lindegren18}. The star has no known binary companions and
is relatively isolated on the sky having just 21 faint sources within $2.5'$
resolved in \gaia{} Data Release 2 \citep[DR2;][]{gaia18}, all of which have
$\Delta G > 6.5$. The astrometric, photometric, and physical stellar parameters
are reported in \autoref{tab:star}.

\begin{deluxetable}{lcc}
\tabletypesize{\small}
\tablecaption{TOI-1235 stellar parameters.\label{tab:star}}
\tablewidth{0pt}
\tablehead{\colhead{Parameter} & \colhead{Value} & \colhead{Refs}}
\startdata 
\multicolumn{3}{c}{\emph{TOI-1235, TIC 103633434, TYC 4384-1735-1,}} \\
\multicolumn{3}{c}{\emph{Gaia DR2 1070387905514406400}} \\
\multicolumn{3}{c}{\emph{Astrometry}} \\
Right ascension (J2015.5), $\alpha$ & 10:08:52.38 & 1,2 \\
Declination (J2015.5), $\delta$ & +69:16:35.83 & 1,2 \\
RA proper motion, $\mu_{\alpha}$ [mas yr$^{-1}$] & $196.63\pm 0.04$ & 1,2 \\
Dec proper motion, $\mu_{\delta}$ [mas yr$^{-1}$] & $17.37\pm 0.05$ & 1,2 \\
Parallax, $\varpi$ [mas] & $25.231\pm 0.030$ & 1,2 \\
Distance, $d$ [pc] & $39.635\pm 0.047$ & 1,2 \\
\multicolumn{3}{c}{\emph{Photometry}} \\
NUV$_{\text{GALEX}}$ & $20.58\pm 0.10$ & 3 \\
$u$ & $15.55\pm 0.30$ & 4 \\
$B_{\text{Tycho-2}}$ & $13.291\pm 0.318$  & 5 \\
$V_{\text{Tycho-2}}$ & $11.703\pm 0.103$  & 5 \\
$V$ & $11.495\pm 0.056$ & 6 \\
$G_{\text{BP}}$ & $11.778\pm 0.002$ & 1,7 \\
$G$ & $10.8492\pm 0.0005$ & 1,7 \\
$G_{\text{RP}}$ & $9.927\pm 0.001$ & 1,7 \\
$T$ & $9.919\pm 0.007$ & 8 \\
$J$ & $8.711\pm 0.020$ & 9 \\
$H$ & $8.074\pm 0.026$ & 9 \\
$K_s$ & $7.893\pm 0.023$ & 9 \\
$W1$ & $7.81\pm 0.03$ & 10 \\
$W2$ & $7.85\pm 0.03$ & 10 \\
$W3$ & $7.77\pm 0.30$ & 10 \\
$W4$ & $7.83\pm 0.22$ & 10 \\
\multicolumn{3}{c}{\emph{Stellar parameters}} \\
$M_V$ & $8.51\pm 0.06$ & 11 \\ 
$M_{K_s}$ & $4.90\pm 0.02$ & 11 \\
Effective temperature, \teff{} [K] & $3872\pm 70$ & 11 \\  
Surface gravity, \logg{} [dex] & $4.646\pm 0.024$ & 11 \\
Metallicity, [Fe/H] & $0.05\pm 0.09$ & 11 \\  
Stellar radius, $R_s$ [R$_{\odot}$] & $0.630\pm 0.015$ & 11 \\ 
Stellar mass, $M_s$ [M$_{\odot}$] & $0.640\pm 0.016$ & 11 \\
Stellar density, $\rho_s$ [g cm$^{-3}$] & $3.61\pm 0.28$ & 11 \\
Stellar luminosity, $L_s$ [L$_{\odot}$] & $0.080\pm 0.007$ & 11 \\
\vspace{-0.15cm} Projected rotation velocity, && \\ \vspace{-0.25cm}
& $<2.6$ & 11 \\
\vsini{} [km s$^{-1}$] && \\
Rotation period, \prot{} [days] & $44.7\pm 4.5$ & 11 \\
\enddata
\tablecomments{\textbf{References:}
  1) \citealt{gaia18}
  2) \citealt{lindegren18}
  3) \citealt{bianchi17}
  4) \citealt{york00}
  5) \citealt{hog00}
  6) \citealt{reid02}
  7) \citealt{evans18}
  8) \citealt{stassun19}
  9) \citealt{cutri03}
  10) \citealt{cutri13}
  11) this work.}
\end{deluxetable}

\begin{figure}
  \centering
  \includegraphics[width=\hsize]{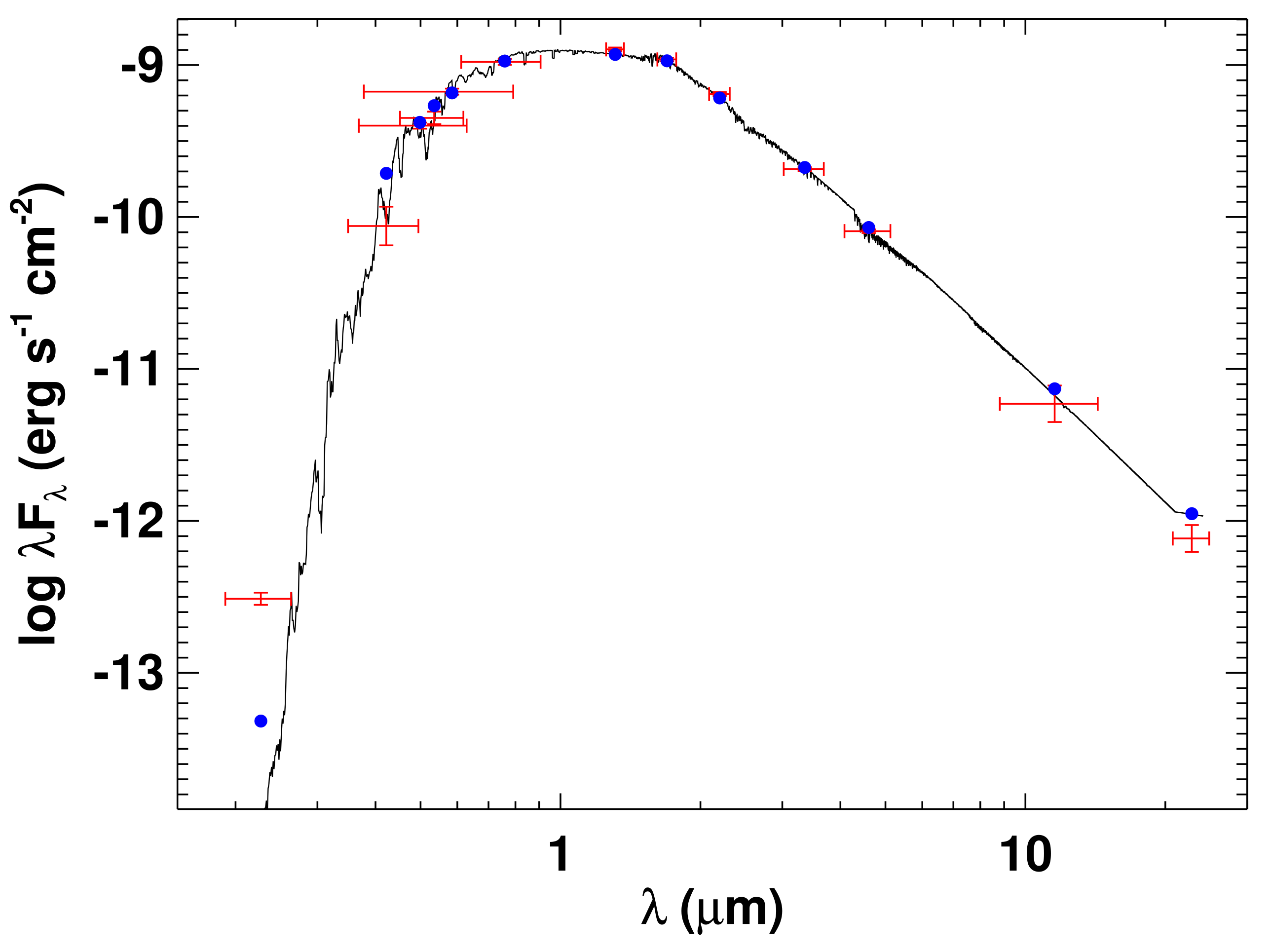}
  \caption{The spectral energy distribution of \name{.} \emph{Red markers}
    depict the photometric measurements with horizontal errorbars depicting
    the effective width of each passband. \emph{Black curve} depicts the most
    likely stellar atmosphere model with \teff{} $=3950$ K.
    \emph{Blue circles} depict the model fluxes over each passband.}
  \label{fig:sed}
\end{figure}

We conducted an analysis of the star's broadband spectral energy distribution
(SED) from the near ultraviolet (NUV) to the mid-infrared (0.23-22 $\mu$m,
\autoref{fig:sed}). We constructed the SED following the procedures outlined
in \cite{stassun16,stassun17b,stassun18b} using retrieved broadband
NUV photometry from \emph{GALEX}, the $u$-band magnitude from the Sloan Digital
Sky Survey, \emph{Tycho-2} $B$ and $V$-band magnitudes, \gaia{} DR2
magnitudes, \emph{2MASS} $JHK_s$ near-IR magnitudes, and \emph{WISE} $W1$-$W4$
IR magnitudes. Assuming zero extinction ($A_V=0$),
we fit the SED with a NextGen stellar atmosphere model \citep{hauschildt99},
treating the metallicity [Fe/H] and effective temperature \teff{} as free
parameters. We derive a weak constraint on [Fe/H] $=-0.5\pm 0.5$ (although we
report the spectroscopically-derived value in \autoref{tab:star}) and measure
\teff{} $=3950\pm 75$ K, which is consistent with \teff{} derived from the
HIRES spectra presented in Sect.~\ref{sect:tks} (\teff{} $=3872\pm 70$ K).
Integrating the SED at a distance of 39.6 pc gives a bolometric flux of
$F_{\text{bol}} = 1.780\pm 0.041 \times 10^{-9}$ erg s$^{-1}$ cm$^{-2}$, which
corresponds to a stellar radius of $0.631\pm 0.024$ \Rsun{.}
As a consistency check, we also fit the SED with a Kurucz stellar atmosphere model 
\citep{kurucz13}. Doing so, we recovered a bolometric flux and stellar radius that
are consistent within $0.5\sigma$ of the values obtained when using the
NextGen stellar models. The inferred stellar radius is also
consistent with the value obtained from the empirically-derived $K_s$-band
radius-luminosity relation from \cite{mann15}: $0.629\pm 0.019$ \Rsun{.} In our
study, we adopt the average of these two values: $R_s=0.630\pm 0.015$ \Rsun{.}
Similarly, we derive the stellar mass using the $K_s$-band mass-luminosity
relation from \cite{benedict16}: $M_s=0.640\pm 0.016$ \Msun{.}

In Sect.~\ref{sect:prot} we report our 
recovery \prot{}$=44.7$ days from archival MEarth photometry. This relatively
long rotation period is consistent with the lack of rotational broadening
observed in our high-resolution spectra presented in Sect.~\ref{sect:rvobs}
(\vsini{} $\leq 2.6$ km s$^{-1}$) and the fact that $H\alpha$ is seen in
absorption (Sect.~\ref{sect:tres}). However, at face value,
the \emph{GALEX} NUV flux in \autoref{fig:sed} appears
to suggest a significant amount of chromospheric emission. This is at odds with
the measured rotation period because, if real,
the apparent excess NUV emission would
imply a Rossby number of 0.2-0.3, or equivalently, \prot{} $=10-15$ days
\citep{stelzer16,wright11}. We note however that the NextGen
atmosphere models do not self-consistently predict M dwarf UV emission from
the chromosphere and transition region such that the apparent NUV excess from
\name{} is unlikely to be a true excess. The absence of chromospheric UV
emission in the atmosphere models is noteworthy as FUV-NUV observations of M
dwarfs have indicated that UV emission is widespread. In other words, even
optically quiescent M dwarfs such as \name{} are known to exhibit
NUV spectra that are qualitatively similar to those of
more active M dwarfs that show chromospheric
$H\alpha$ in emission \citep{walkowicz08,france13}.
Furthermore, the empirical \texttt{GALEX} NUV-$K_s$ color
relation with NUV flux from \cite{ansdell15}, derived the early M dwarf observations,
reveals that $\log{F_{\text{NUV}}/F_{\text{bol}}}=-4.7\pm 0.1$ for TOI-1235. This value
is significantly less than $\log{F_{\text{NUV}}/F_{\text{bol}}}=-3.8\pm 0.1$ based on the
stellar atmosphere models used here. This discrepancy between observations of early M
dwarfs and models supports the notion that the
apparent NUV excess exhibited in \autoref{fig:sed} is not a true NUV excess.

\section{Observations} \label{sect:observations}
\subsection{TESS photometry} \label{sect:tessphot}
\name{} was observed in three non-consecutive \tess{} sectors between UT July
18, 2019 and February 18, 2020. \name{} is a member of the Cool
Dwarf target list \citep{muirhead18} and was included in the \tess{}
Input Catalog \citep[TIC;][]{stassun17a}, the \tess{} Candidate Target List
(CTL), and in the Guest
Investigator program 22198\footnote{PI: Courtney Dressing.}, such that its light
curve was sampled at 2-minute cadence. \tess{} observations occurred in CCD 3 on
Camera 4 in Sector 14 (UT July 18-August 14 2019), in CCD 1 on Camera 2 in
Sector 20 (UT December 24 2019-January 20 2020), and in CCD 2 on Camera 2 in
Sector 21 (UT January 21-February 18 2020). Sector 14 was the first pointing
of the spacecraft in the northern ecliptic hemisphere. As indicated in the
data release notes\footnote{\url{https://archive.stsci.edu/tess/tess_drn.html}},
to avoid significant
contamination in cameras 1 and 2 due to scattered light by the Earth and Moon,
the Sector 14 field was pointed to +85$^{\circ}$ in ecliptic latitude,
31$^{\circ}$ north of its intended pointing from the nominal mission strategy.
Despite this, all cameras in Sector 14 continued to be affected by scattered
light for longer periods of time compared to most other sectors due to the
Earth's position above the sunshade throughout the orbit.
Camera 2 during sectors 20 and 21 was largely unaffected by scattered light
except during data downloads and at the beginning of the second orbit in Sector
21 due to excess Moon glint.

\begin{figure*}
  \centering
  \includegraphics[width=\hsize]{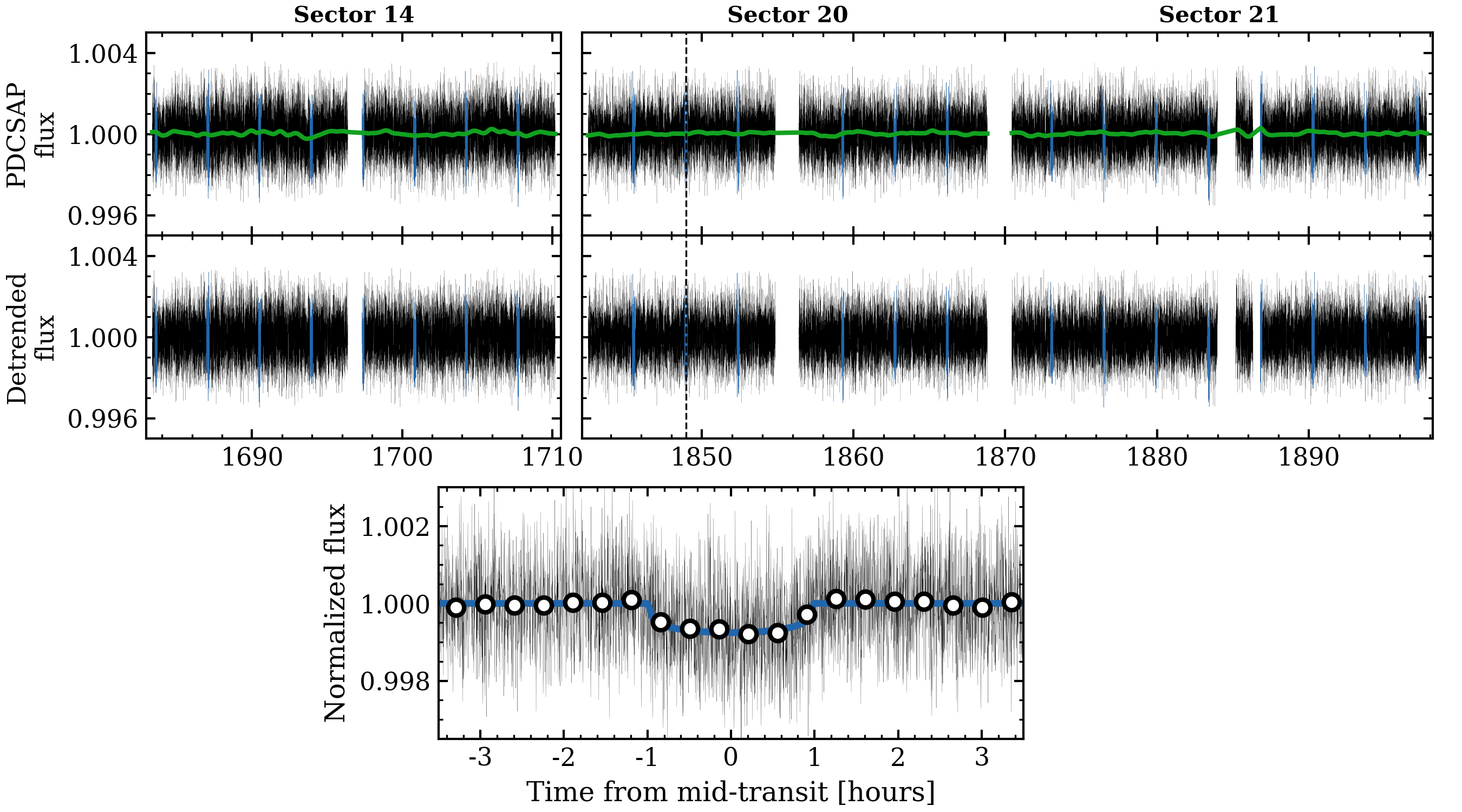}
  \caption{\tess{} light curve of \name{} from sectors 14, 20, and 21.
    \emph{Upper row}: the \texttt{PDCSAP} light curve following the removal of
    systematics via a linear combination of Cotrending Basis Vectors. The
    \emph{green curve} depicts the mean GP model of residual temporally
    correlated noise (Sect.~\ref{sect:tessmcmc}). The $3\sigma$ uncertainties
    on the mean GP model are smaller than the curve width. 
    In-transit measurements are highlighted in blue throughout. The
    \emph{vertical dashed line} highlights the epoch of the ground-based transit
    observation from LCOGT, which confirms the transit event on-target
    (Sect.~\ref{sect:sg1}). \emph{Lower row}: the detrended \texttt{PDCSAP}
    light curve. \emph{Bottom panel}: the phase-folded transit light curve
    of \name{} b from 22 individual transit events. The maximum a-posteriori
    transit model is depicted by the \emph{blue curve} while the
    \emph{white markers} depict the binned photometry.}
  \label{fig:tess}
\end{figure*}

The \tess{} images were
processed by the NASA Ames Science Processing Operations Center
\citep[SPOC;][]{jenkins16}, which produce two light curves per sector called
Simple Aperture Photometry (\texttt{SAP}) and Presearch Data Conditioning Simple
Aperture Photometry \citep[\texttt{PDCSAP};][]{smith12,stumpe12,stumpe14}.
The light curves are
corrected for dilution during the SPOC processing with \name{} suffering only
marginal contamination with a dilution correction factor of 0.9991.
Throughout, we only consider reliable \tess{} measurements for which
the measurement's quality flag \texttt{QUALITY} is equal to zero.
The \texttt{PDCSAP} light curve is constructed by detrending the \texttt{SAP}
light curve using a linear combination of Cotrending Basis Vectors
(CBVs), which are derived from a principal component decomposition of the light
curves on a per sector per camera per CCD basis. \name{'}s \texttt{PDCSAP}
light curve is depicted in \autoref{fig:tess} and shows no compelling signs of
coherent photometric variability from rotation. However,
the set of CBVs (not shown) exhibit sufficient temporal structure such that a
linear combination of CBVs can
effectively mask stellar rotation signatures greater than a few days. Thus,
inferring \prot{} for \name{} from \tess{} would be challenging and is
addressed more effectively with ground-based photometric monitoring in
Sect.~\ref{sect:prot}.

Following light curve construction, the SPOC conducts a subsequent transit
search on each sector's \texttt{PDCSAP}
light curve using the Transiting Planet Search Module
\citep[TPS;][]{jenkins02,jenkins10}. The TOI-1235.01 transit-like signal was
detected in all three sectors independently and passed a set of internal data
validation tests \citep{twicken18,li19}. The reported period of the planet
candidate was 3.44 days in Sectors 14 and 20 and three times that value (i.e.
10.33 days) in Sector 21 due to the low signal-noise ratio of the individual
transits. At 3.44 days, there are eight, six, and eight transits observed in
each of the three sectors. The transit events are highlighted in
\autoref{fig:tess}. The SPOC reported a preliminary transit depth
of $841\pm 72$ ppm, which corresponded to a planetary radius of
$2.0\pm 0.1$ \Rearth{} using our stellar radius (\autoref{tab:star}).

\subsection{Photometric monitoring with MEarth} \label{sect:prot}

\begin{figure*}
  \centering
  \includegraphics[width=\hsize]{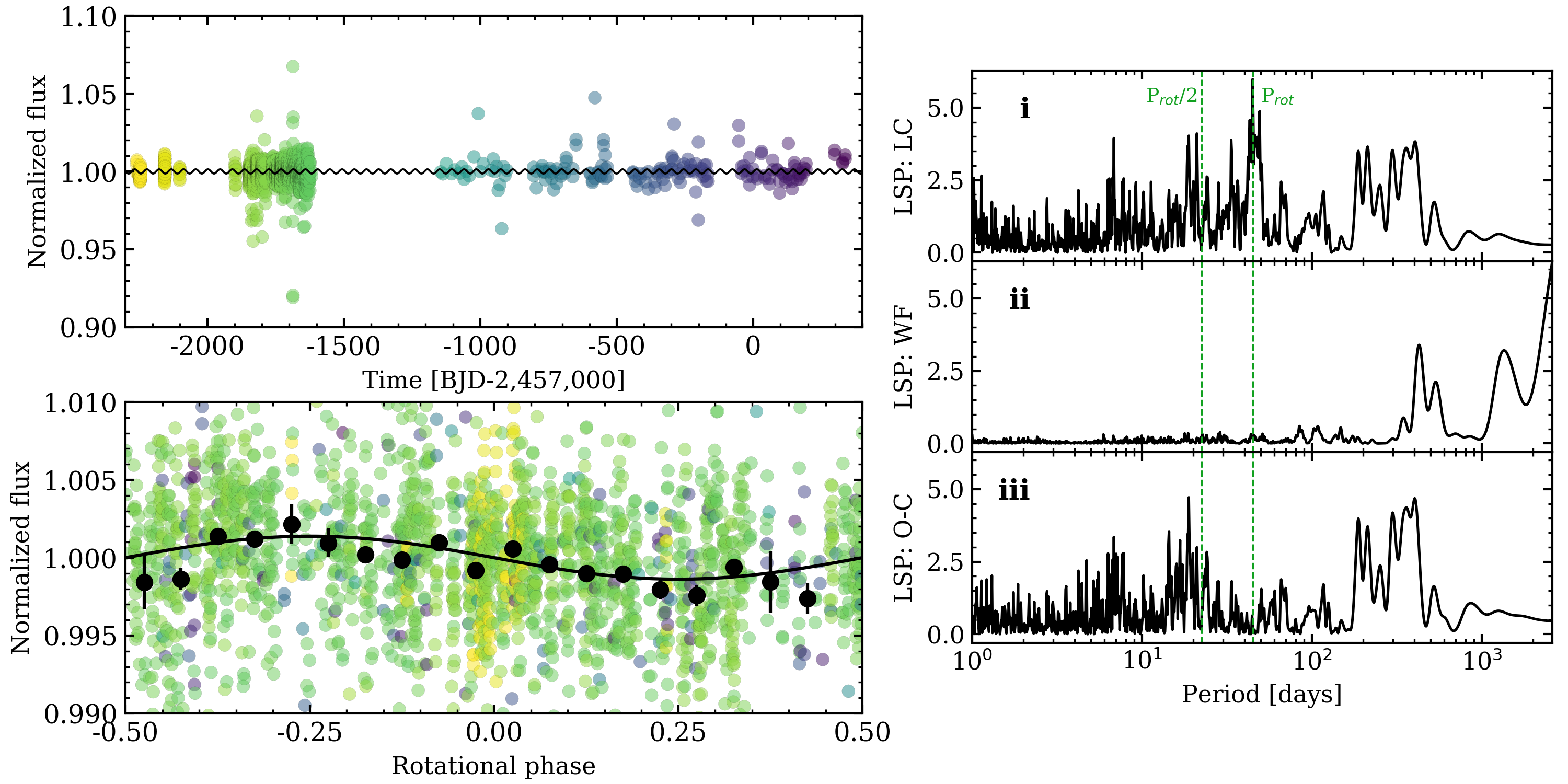}
  \caption{Measurement of the \name{} photometric rotation period with
    MEarth-North. \emph{Upper left panel}: the \name{} differential light curve
    from archival MEarth-North photometry (October 2008-November 2015).
    \emph{Right panels}: Lomb-Scargle periodograms of i) the detrended
    light curve, ii) the window function, and iii) the photometric residuals
    after removal of the optimized sinusoidal fit with \prot{}$=44.7$ days.
    \emph{Lower left panel}: the light curve phase-folded to \prot{.}
    \emph{Circular black markers} represent the binned light curve while the
    \emph{solid black curve} depicts the sinusoidal fit.}  
  \label{fig:prot}
\end{figure*}

Inactive early M dwarfs have typical rotation periods of 10-50 days
\citep{newton17}. In Sect.~\ref{sect:tessphot} we described how measuring
\prot{} for \name{} with \tess{} is intractable due to the flexibility in the
systematics model. Fortunately, MEarth-North has archival images of the field
surrounding \name{} that span 7.1 years (UT October 2, 2008 to November 10,
2015) from which \prot{} may be measured.
MEarth-North is a telescope array located at the Fred Lawrence Whipple
Observatory
(FLWO) on Mount Hopkins, AZ. The facility consists of eight 40cm telescopes,
each equipped with a $25.6' \times 25.6'$ field-of-view Apogee U42 camera, with
a custom passband centered in the red optical (i.e. RG715). MEarth-North has
been photometrically monitoring nearby mid-to-late M dwarfs ($<0.33 $\Rsun{)}
since 2008, in search of transiting planets \citep{berta12,irwin15} and
to conduct detailed studies of stellar variability \citep{newton16a}. Although
\name{} was too large to be included in the initial target list
\citep{nutzman08}, its position happens to be within $14'$ of an intentional
target (GJ 1131) such that we are able to construct and analyze its light curve
here for the first time.

To search for photometric signatures of rotation, we first
retrieved the archival image sequence and computed the
differential light curve of \name{} as shown in \autoref{fig:prot}. We then
investigated the Lomb-Scargle periodogram (LSP) of the light curve, which
reveals a significant peak around 45 days that
is not visible in the LSP of the window function (\autoref{fig:prot}).
Using this value as an initial guess, we proceeded with
fitting the light curve following the methods outlined
in \cite{irwin06,irwin11}. The model includes systematics terms,
predominantly from variations
in the precipitable water vapor (PWV) column above the telescope,
plus a sinusoidal term to model rotational modulation.
As outlined in \citep{newton16a}, a ``common mode'' vector is constructed
as a low cadence comparison light curve that tracks variations in the PWV
and is included in our systematics model as a linear term along with the
full width at half maximum of the MEarth point spread function. With
this full model, we measure \prot{}
$=44.7\pm 4.5$ days and a variability semi-amplitude of 1.33 ppt.
The detrended light curve, phase-folded to \prot{,} is included in
\autoref{fig:prot}. \autoref{fig:prot} also reveals that the subtraction of
our systematics plus rotation model from the light curve, mitigates the
45-day signal in the LSP with no significant residual periodicities.
The shallow
variability amplitude is unsurprising for relatively warm early M dwarfs like
\name{} whose spot-to-photosphere temperature contrasts are small
\citep{newton16a}. 
We note that knowledge
of \prot{} can be critical for the interpretation of RV signals as even active
regions with small temperature contrasts
can induce large RV variations due to the suppression of
convective blueshift \citep{dumusque14}.

\subsection{Reconnaissance spectroscopy with TRES} \label{sect:tres}

\begin{figure*}
  \centering
  \includegraphics[width=\hsize]{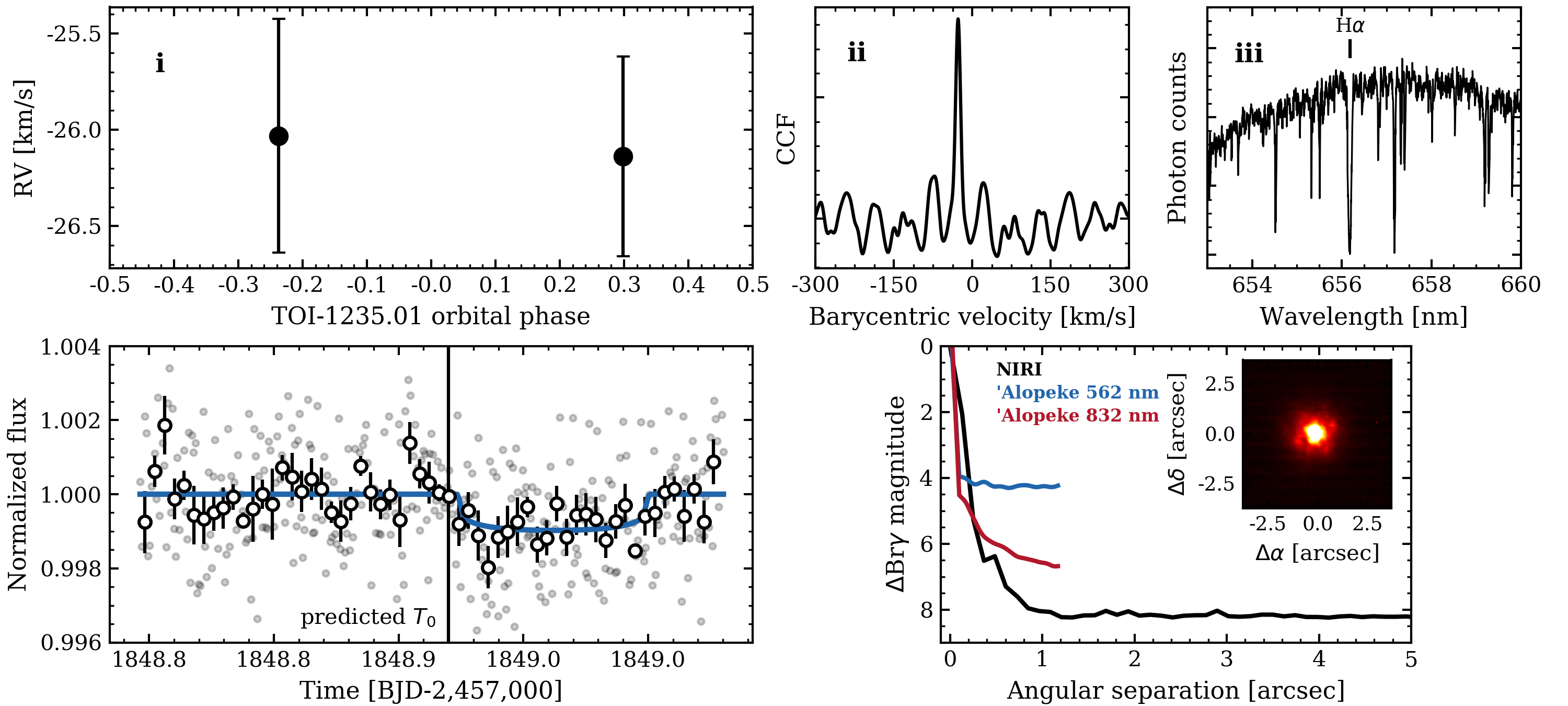}
  \caption{Summary of TFOP follow-up observations of \name{} for planet
    validation
    purposes. \emph{Top row}: results from TRES reconnaissance spectroscopy
    that i) show no RV variations thus ruling out a spectroscopic binary, ii)
    reveal a single-lined CCF with no rotational broadening, and iii) show
    $H\alpha$ in absorption. \emph{Lower left panel}: the ground-based transit
    light curve obtained with LCOGT showing that the expected transit event
    occurred on-target and arrived 63 minutes late relative to the SPOC-reported
    linear ephemeris represented by the \emph{black vertical line}.
    \emph{Open circles} depict the light curve in 5.5 minute bins. The
    \emph{blue curve} depicts the optimized transit fit to the LCOGT photometry.
    \emph{Lower right panel}: the $5\sigma$ contrast curves from
    \emph{Gemini}/NIRI AO-imaging (\emph{black}), `Alopeke 562 nm speckle imaging
    (\emph{blue}), and `Alopeke 832 nm speckle imaging (\emph{red}).
    The inset depicts the coadded image from \emph{Gemini}/NIRI
    AO-imaging centered on \name{.}}
  \label{fig:tfop}
\end{figure*}

We began to pursue the confirmation of the planet candidate TOI-1235.01 by
obtaining reconnaissance spectra with the Tillinghast Reflector
\'Echelle Spectrograph (TRES) through coordination with the \tess{} Follow-up
Observing Program (TFOP). TRES is a fiber-fed $R=44,000$ optical \'echelle
spectrograph (310-910 nm), mounted on the 1.5m Tillinghast Reflector
telescope at FLWO. Multiple spectra were obtained to search for radial
velocity (RV) variations indicative of a spectroscopic binary, and to assess
the level of surface rotation and chromospheric activity. We obtained two
spectra at opposite quadrature phases of TOI-1235.01 on UT December 1 and 13,
2019 with exposure times of 2100 and 1200 seconds, which resulted in a S/N per
resolution element of 31.4 and 26.0 respectively at 519 nm in the order
containing the information-rich Mg b lines.

The TRES RVs phase-folded to the TOI-1235.01 ephemeris are depicted in
\autoref{fig:tfop} and show no significant variation thus
ruling out a spectroscopic binary. The cross-correlation function of the median
spectrum with a rotating template of Barnard's star is also shown in
\autoref{fig:tfop} and reveals a single-lined spectrum with no significant
rotational broadening (\vsini{} $<3.4$ km s$^{-1}$). Lastly,
the $H\alpha$ feature shown is seen in absorption, which is indicative of a
chromospherically inactive star and is consistent with \prot{}$\gtrsim 10$ days
\citep{newton17}. Taken together, our reconnaissance spectra
maintain that TOI-1235.01 is a planetary candidate around a relatively inactive
star.

\subsection{Ground-based transit photometry with LCOGT} \label{sect:sg1}
\tess{'}s large pixels ($21''$) can result in significant blending of target
light curves with nearby sources. To confirm that the transit event occurs
on-target, and to rule out nearby eclipsing binaries (EBs), we targeted a
transit of TOI-1235.01 with seeing-limited photometric follow-up on UT
December 31, 2019. This observation was scheduled after the planet candidate was
detected in \tess{} Sector 14 only and occurred during Sector 20. The
transit observation was scheduled using the \tess{} Transit Finder, a customized
version of the \texttt{Tapir} software package \citep{jensen13}. We
obtained a $z_s$-band light curve from the McDonald Observatory with the 1-meter
telescope as part of the Las Cumbres Observatory Global Telescope network
\citep[LCOGT;][]{brown13}. The telescope is equipped with a $4096\times 4096$
Sinistro camera whose pixel scale is 54 times finer than that of \tess{}:
0.389 $''$ pixel$^{-1}$. We calibrated the full image sequence using the
standard LCOGT
\texttt{BANZAI} pipeline \citep{mccully18}. The differential photometric light
curve of \name{,} along with seven sources within $2.5'$, were derived
from $7''$ uncontaminated
apertures using the \texttt{AstroImageJ} software package
\citep[AIJ;][]{collins17}. The field was cleared of nearby EBs down to
$\Delta z_s = 7.15$ as we did not detect eclipses from neighboring sources
close to the expected transit time.

A full transit event was detected on-target and is included in
\autoref{fig:tfop}. We fit the light curve with a
\cite{mandel02} transit model calculated using the \texttt{batman} software
package \citep{kreidberg15}. The shallow transit depth of TOI-1235.01 produces
a low S/N transit that does not provide strong constraints on most model
parameters relative to what can be recovered from 22 transits in \tess{.}
Consequently, we fix the model to a circular orbit with an orbital period,
scaled semimajor axis, and impact parameter of $P=3.44471$ days, $a/R_s=13.2$,
and $b=0.45$ respectively. Furthermore, we set the quadratic limb-darkening
parameters in the $z_s$-band to $u_1=0.25$ and $u_2=0.33$ as interpolated from
the \cite{claret11} tables using the \texttt{EXOFAST} tool \citep{eastman13}. 
We fit the baseline flux, time of mid-transit, and planet-star
radius ratio via non-linear least squares optimization using the
\texttt{scipy.curve\_fit} function and find that $f_0=1.000$,
$T_0=2,458,848.962$ BJD, and $r_p/R_s=0.0295$.
The transit is seen to arrive 63 minutes late relative to the
linear ephemeris reported by the SPOC from Sector 14 only. 
The transit depth of 0.867 ppt is
$4.5\sigma$ deeper than the \tess{} transit measured in our fiducial analysis
(0.645 ppt, Sect.~\ref{sect:tessmcmc}). Due to the similar wavelength coverage
between the $z_s$ and \tess{} passbands, and because of the large residual
systematics often suffered by ground-based light curves of shallow transits,
we attribute this discrepancy to unmodeled systematics rather than to a
bona-fide chromatic transit depth variation.

\subsection{High resolution imaging}
\tess{'}s large pixels also make the \tess{} light curves susceptible to
contamination by very nearby sources that are not detected in \gaia{} DR2, nor
in the seeing-limited image sequences. To clear the field of very nearby sources
and a possible false positive in the form of a blended EB \citep{ciardi15}, we
obtained two independent sets of high-resolution follow-up
imaging sequences as described in the following sections.

\subsubsection{Adaptive optics imaging with Gemini/NIRI}
We obtained
adaptive-optics (AO) images with \emph{Gemini}/NIRI \citep{hodapp03}
on UT November 25, 2019 in the Br$\gamma$ filter. We collected 9
dithered images with integration times of 3.5 seconds. The data were
reduced following a standard reduction procedure that includes bad pixel
corrections, flat-fielding, sky subtraction, and image coaddition.
The $5\sigma$ contrast curve and the coadded image of \name{} are
included in the lower right panel of \autoref{fig:tfop}. These data provide
sensitivity to visual companions with $\Delta \text{Br}\gamma \leq 5$ for
separations $> 270$ mas and $\Delta \text{Br}\gamma \leq 8.2$ beyond
$1''$. We do not detect any visual companions within $5''$ of \name{}
within the $5\sigma$ sensitivity of our observations.

\subsubsection{Speckle imaging with Gemini/`Alopeke}
We also obtained speckle interferometric images on UT February 16, 2020
using the `Alopeke
instrument\footnote{\url{https://www.gemini.edu/sciops/instruments/alopeke-zorro/}}
mounted on the 8-meter Gemini North telescope on the summit of Mauna Kea in Hawai'i. 
`Alopeke simultaneously observes diffraction-limited images at 562 nm and 832
nm. Our data set consisted of 3 minutes of total integration time taken as sets of
$1000 \times 0.06$ second images. Following \citep{howell11}, we combined all images
subjected to Fourier analysis to produce the speckle reconstructed imagery from which
the $5\sigma$ contrast curves are derived in each passband (lower right panel of
\autoref{fig:tfop}). Our data reveal \name{} to be a single star to contrast limits
of 4.5 to 7 magnitudes, eliminating essentially all main sequence stars fainter than
\name{} within the spatial limits of 0.8 to 48 AU.

Using our reconnaissance spectroscopy, ground-based transit follow-up, and
high resolution imaging observations as input (\autoref{fig:tfop}), we used the
\texttt{vespa} and \texttt{triceratops}
statistical validation tools to compute the TOI-1235.01 false
positive probability (FPP) \citep{morton12,giacalone20}.
In both analyses we find that FPP $<1$\% and will refer to the 
validated planet as \name{} b for the remainder of this study.

\subsection{Precise radial-velocities} \label{sect:rvobs}
\subsubsection{HARPS-N}
We obtained 27 spectra of \name{} with the HARPS-N optical \'echelle
spectrograph at the 3.6m Telescopio Nazionale Galileo on La Palma in the Canary
Islands. The HARPS-N optical spectrograph, with a resolving power of
$R = 115,000$, is stabilized in
pressure and temperature, which enable it to achieve sub-\mps{} accuracy under
ideal observing conditions when sufficient S/N is attainable
\citep{cosentino12}. The spectra
were taken as part of the HARPS-N Collaboration Guaranteed Time Observations
program between UT December 24, 2019 and March 12, 2020. The exposure time was
set to 1800 seconds. In orders redward of order 18 (440-687 nm), we achieved
a median S/N of 45.2 and a median measurement uncertainty of 1.22
\mps{.}
\name{} did not exhibit any rotational broadening in the HARPS-N
spectra leading to \vsini{} $\leq 2.6$ km s$^{-1}$, a result that is consistent
with its measured rotation period \prot{} $=44.7\pm 4.5$ days.

We extracted the HARPS-N RVs using the \texttt{TERRA} pipeline
\citep{anglada12}. \texttt{TERRA} employs a template-matching scheme that is
known to achieve improved RV measurement uncertainties on M dwarfs relative to
the cross-correlation function (CCF) technique \citep{anglada12}. M dwarfs are
particularly well-suited to RV extraction via template-matching because the
line lists used to define the binary mask for the CCF technique are
incomplete and often produce a CCF template that is a poor match for cool M
dwarfs. A master template
spectrum is constructed by first shifting the individual spectra to the
barycentric frame using the barycentric corrections calculated by the HARPS-N
Data Reduction Software (DRS; \citealt{lovis07}), after masking portions of the
wavelength-calibrated spectra wherein telluric absorption is $\geq 1$\%. A
high S/N template spectrum is then built by coadding the individual
spectra. \texttt{TERRA} then computes the RV of each spectrum relative to the
template via least-squares matching the spectrum in velocity space. Throughout
the extraction process, we only consider orders redward of order 18 
such that the bluest orders at low S/N are ignored. The resulting RV time series
is provided in \autoref{tab:rvs}.

\begin{deluxetable}{cccc}
\tablecaption{Radial velocity time series of \name{} from HARPS-N \& HIRES\label{tab:rvs}}
\tablewidth{0pt}
\tablehead{Time & RV & $\sigma_{\text{RV}}$ & Instrument \\
$[$BJD - 2,457,000$]$ & $[\text{m s}^{-1}]$ & $[\text{m s}^{-1}]$ & }
\startdata
1890.653258 & -0.119 & 0.975 & HARPS-N \\
1905.851683 & -7.358 & 1.281 & HIRES \\
1906.724763 & 1.803 & 1.470 & HARPS-N \\
\enddata
\tablecomments{For conciseness, only a subset of three rows are depicted here to illustrate the table's contents. The entirety of this table is provided in the arXiv source code.}
\end{deluxetable}

\subsubsection{HIRES} \label{sect:tks}
We obtained 11 additional spectra of \name{} with the High Resolution \'Echelle
Spectrometer on Keck-I \citep[HIRES;][]{vogt94} as part of the \tess{-}Keck
Survey (TKS) between UT December 10, 2019 and March 10, 2020. HIRES
is an optical spectrograph at $R=60,000$ that uses a
heated iodine cell in front of the spectrometer entrance slit to perform its
precise wavelength calibration between 500-620 nm. Against the forest of iodine
cell features imprinted on the spectrum, we measure the relative Doppler shift
of each spectrum while constraining the shape of the instrument profile at each
epoch \citep{howard10}. The median exposure time was set to 
900 seconds, which resulted in a median
S/N at 550 nm of 124 and a median measurement uncertainty of 1.21 \mps{,} nearly
identical to the median RV uncertainty in our HARPS-N time series. The HIRES
RV measurements are also provided in \autoref{tab:rvs}.

We processed a single epoch spectrum with a S/N of 96 per pixel
using the \texttt{SpecMatch-Emp} algorithm
\citep{yee17} to independently derive spectroscopic stellar parameters. The
resulting effective temperature and metallicity are reported in
\autoref{tab:star}. We also infer a stellar radius of $R_s=0.61\pm 0.10$
\Rsun{,} which is consistent with the values derived from our SED analysis and
from the empirical M dwarf radius-luminosity relation.

\section{Data Analysis \& Results} \label{sect:analysis}
Here we conduct a pair of independent analyses of our data to test the
robustness of the recovered planetary parameters following the strategy
adopted in \cite{cloutier20b}. In our fiducial analysis
(Sects.~\ref{sect:tessmcmc} and~\ref{sect:rvs}), we model the \tess{} light
curve independently and use the resulting planet parameter posteriors as priors
in our subsequent RV analysis. In Sect.~\ref{sect:exofast} we conduct an
alternative global analysis using the \texttt{EXOFASTv2} software
\citep{eastman19}.

\subsection{TESS transit analysis} \label{sect:tessmcmc}
We begin our fiducial analysis by modeling the \tess{} \texttt{PDCSAP} light
curve (\autoref{fig:tess}) in which the planet candidate TOI-1235.01 was
originally detected. The \texttt{PDCSAP} light curve
has already undergone systematics corrections via a linear combination of
CBVs however, some low amplitude temporally-correlated signals that are
unrelated to planetary transits are seen to persist. We elect to model these
signals as an untrained semi-parametric Gaussian process (GP)
simultaneously with the transit model of \name{} b.
We employ the \texttt{exoplanet} software
package \citep{foremanmackey19} to construct the GP and transit model in each
step in our Markov Chain Monte-Carlo (MCMC) simulation. Within
\texttt{exoplanet}, analytical transit models are computed using the
\texttt{STARRY} package \citep{luger19} while \texttt{celerite}
\citep{foremanmackey17} is used to evaluate the marginalized likelihood of the
GP model.

We adopt a covariance kernel of the form of a stochastically-driven
simple harmonic oscillator in Fourier space. The power spectral density of
the kernel is

\begin{equation}
  S(\omega) = \sqrt{\frac{2}{\pi}} \frac{S_0 \omega_0^4}{(\omega^2-\omega_0^2)^2 + \omega_0^2 \omega^2 / Q^2},
\end{equation}

\noindent which is parameterized by the frequency of the undamped oscillator
$\omega_0$, the factor $S_0$, which is proportional to the spectral power at
$\omega_0$, and the fixed quality factor $Q=\sqrt{0.5}$. We also include
the baseline flux $f_0$ and an additive scalar jitter $s_{\text{TESS}}$ in our 
noise model that we parameterize by
$\{ \ln{\omega_0}, \ln{S_0 \omega_0^4}, f_0, \ln{s_{\text{TESS}}^2} \}$. Our noise
model is jointly fit with a transit model for \name{} b with the following
free parameters: the stellar mass $M_s$, stellar radius $R_s$, quadratic
limb-darkening coefficients
$\{ u_1, u_2 \}$, orbital period $P$, time of mid-transit $T_0$, planet radius
$r_p$, impact parameter $b$, eccentricity $e$, and argument of periastron
$\omega$. Our full \tess{} model therefore contains 13 free parameters that are
parameterized by $\{ f_0, \ln{\omega_0}, \ln{S_0 \omega_0^4},
\ln{s_{\text{TESS}}^2}, M_s,R_s,u_1,u_2,\ln{P}, T_0, \ln{r_p},$
$b, e, \omega \}$.
Our adopted model parameter priors are listed in \autoref{tab:priors}.

\begin{deluxetable*}{lcc}
\tabletypesize{\small}
\tablecaption{\tess{} light curve and RV model parameter priors\label{tab:priors}}
\tablewidth{0pt}
\tablehead{Parameter & Fiducial Model Priors & \texttt{EXOFASTv2} Model Priors}
\startdata
\multicolumn{3}{c}{\emph{Stellar parameters}} \\
\teff{,} [K] & $\mathcal{N}(3872,70)$ & $\mathcal{N}(3872,70)$ \\
$M_s$, [\Msun{]} & $\mathcal{N}(0.640,0.016)$ & $\mathcal{N}(0.640,0.016)$ \\
$R_s$, [\Rsun{]} & $\mathcal{N}(0.630,0.015)$ & $\mathcal{N}(0.630,0.015)$ \\
\multicolumn{3}{c}{\emph{Light curve hyperparameters}} \\
$f_0$ & $\mathcal{N}(0,10)$ & $\mathcal{U}(-\inf,\inf)$ \\ 
$\ln{\omega_0}$, [days$^{-1}$] & $\mathcal{N}(0,10)$ & - \\
$\ln{S_0\omega_0^4}$ & $\mathcal{N}(\ln{\text{var}(\mathbf{f}_{\text{TESS}})},10)$ & - \\
$\ln{s_{\text{TESS}}^2}$ & $\mathcal{N}(\ln{\text{var}(\mathbf{f}_{\text{TESS}})},10)$ & - \\
$u_1$ & $\mathcal{U}(0,1)$ & $\mathcal{U}(0,1)$ \\
$u_2$ & $\mathcal{U}(0,1)$ & $\mathcal{U}(0,1)$ \\
Dilution & - & $\mathcal{N}(0,0.1\: \delta)$\tablenotemark{a} \\ 
\multicolumn{3}{c}{\emph{RV parameters}} \\
$\ln{\lambda}$, [days] & $\mathcal{U}(\ln{1},\ln{1000})$ & - \\
$\ln{\Gamma}$ & $\mathcal{U}(-3,3)$ & -  \\
$P_{\text{rot}}$, [days] & $\mathcal{N}(46.1,4.6)$ & -  \\
$\ln{a}_{\text{HARPS-N}}$, [\mps{]} & $\mathcal{U}(-5,5)$ & -  \\
$\ln{a}_{\text{HIRES}}$, [\mps{]} & $\mathcal{U}(-5,5)$ & -  \\
$\ln{s}_{\text{HARPS-N}}$, [\mps{]} & $\mathcal{U}(-5,5)$ & $\mathcal{U}(-\inf,\inf)$ \\
$\ln{s}_{\text{HIRES}}$, [\mps{]} & $\mathcal{U}(-5,5)$ & $\mathcal{U}(-\inf,\inf)$ \\
$\gamma_{\text{HARPS-N}}$, [\mps{]} & $\mathcal{U}(-10,10)$ & $\mathcal{U}(-\inf,\inf)$ \\
$\gamma_{\text{HIRES}}$, [\mps{]} & $\mathcal{U}(-10,10)$ & $\mathcal{U}(-\inf,\inf)$ \\
\multicolumn{3}{c}{\emph{\name{} b parameters}} \\
$P$, [days] & $\mathcal{U}(-\inf,\inf)$\tablenotemark{b} & $\mathcal{U}(-\inf,\inf)$ \\
$T_{0}$, [BJD-2,457,000] & $\mathcal{U}(-\inf,\inf)$\tablenotemark{b} & $\mathcal{U}(-\inf,\inf)$ \\
$\ln{r_{p}}$, [\Rearth{]} & $\mathcal{N}(0.5\cdot \ln(Z) + \ln{R_s},1)$\tablenotemark{c} & -\\
$r_{p}/R_s$ & - & $\mathcal{U}(-\inf,\inf)$ \\
$b$ & $\mathcal{U}(0,1+r_{p,b}/R_s)$ & - \\
$\ln{K}$, [\mps{]} & $\mathcal{U}(-5,5)$ & - \\
$K$, [\mps{]} & - & $\mathcal{U}(-\inf,\inf)$ \\
$e$ & $\mathcal{B}(0.867,3.03)$\tablenotemark{d} \\
$\omega$, [rad] & $\mathcal{U}(-\pi,\pi)$ \\
$e\cos{\omega}$ & - & $\mathcal{U}(-1,1)$ \\
$e\sin{\omega}$ & - & $\mathcal{U}(-1,1)$ \\
$\sqrt{e}\cos{\omega}$ & $\mathcal{U}(-1,1)$ - & \\
$\sqrt{e}\sin{\omega}$ & $\mathcal{U}(-1,1)$ - & \\
\enddata
\tablecomments{Gaussian distributions are denoted by $\mathcal{N}$ and are
  parameterized by mean and standard deviation values. Uniform distributions
  are denoted by $\mathcal{U}$ and bounded by the specified lower and upper
  limits. Beta distributions are denoted by $\mathcal{B}$ and are parameterized
  by the shape parameters $\alpha$ and $\beta$.}
\tablenotetext{a}{$\delta$ is the SPOC-derived dilution factor applied to the \tess{} light curve.}
\tablenotetext{b}{This prior in the fiducial model reflects that used in the \tess{} light analysis. However, its resulting posterior is used as an informative prior in the subsequent RV analysis.}
\tablenotetext{c}{The transit depth of TOI-1235.01 reported by the SPOC: $Z=841$ ppm.}
\tablenotetext{d}{\citealt{kipping13}.}
\end{deluxetable*}

We execute an MCMC to sample the joint posterior probability density 
function (PDF) of our full set of model parameters using the \texttt{PyMC3} MCMC
package \citep{salvatier16} within \texttt{exoplanet}. The MCMC is initialized
with four simultaneous chains, each with 4000 tuning steps and 3000 draws in the
final sample. Point estimates of the maximum a-posteriori (MAP) values from the
marginalized posterior PDFs of the GP hyperparameters are selected to construct
the GP predictive
distribution whose mean function is treated as our detrending model of the
\texttt{PDCSAP} light curve. This mean detrending function and the detrended
light curve are both shown in \autoref{fig:tess}.
Similarly, we recover the MAP point estimates of the transit model parameters
to construct the transit model shown in the bottom panel of \autoref{fig:tess}.
MAP values and uncertainty point estimates from the $16^{\text{th}}$ and
$84^{\text{th}}$ 
percentiles for all model parameters are reported in \autoref{tab:results}.

\subsection{Precise radial-velocity analysis} \label{sect:rvs}
We continue our fiducial analysis by jointly modeling the HARPS-N and HIRES
RV time series. Here we are able to exploit the strong priors on $P$ and $T_0$
derived from our analysis of the \tess{} light curve
(Sect.~\ref{sect:tessmcmc}).

\begin{figure*}
  \centering
  \includegraphics[width=\hsize]{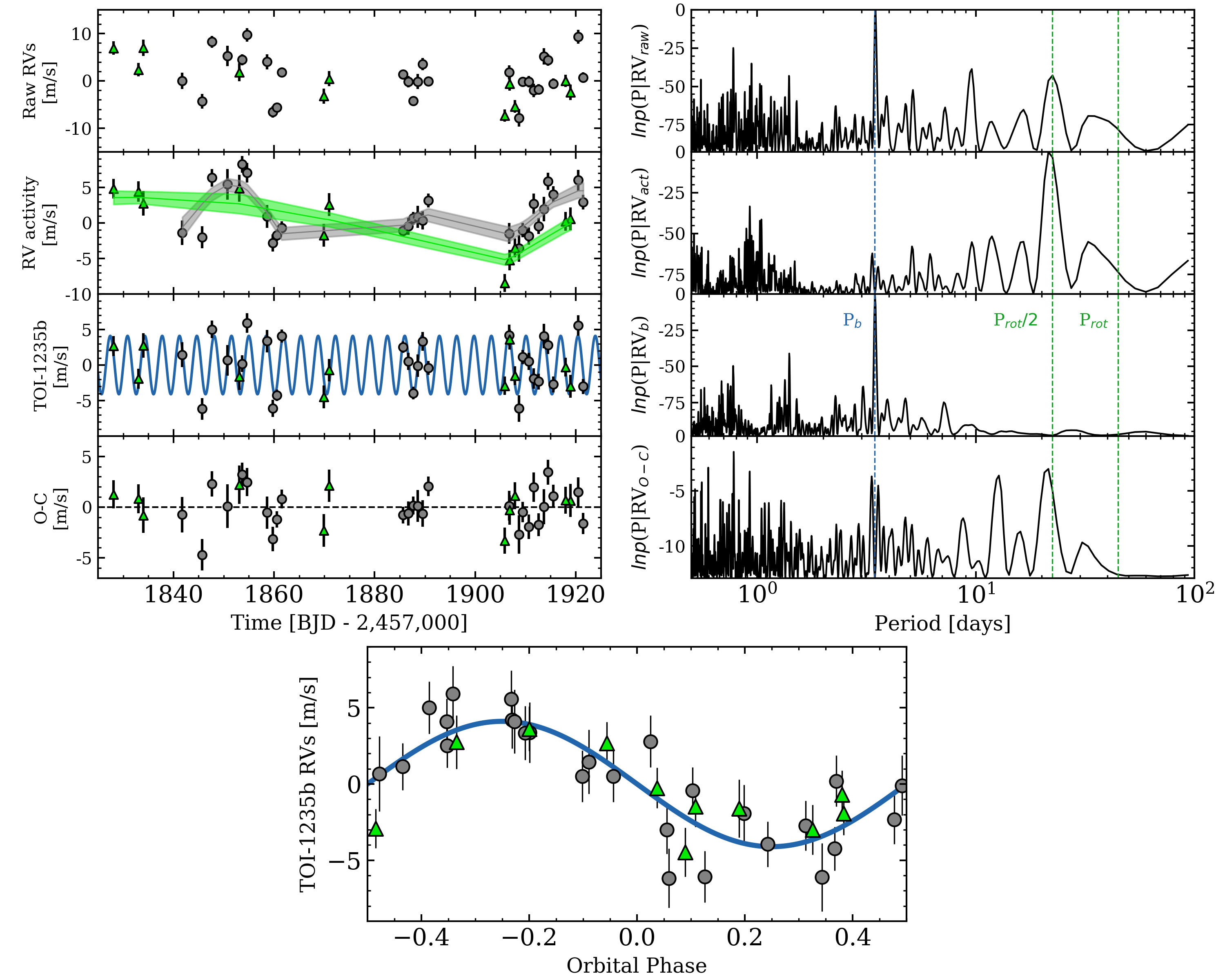}
  \caption{\name{} RVs from HARPS-N (\emph{gray circles}) and HIRES
    (\emph{green triangles}). The data of each RV component and 
    its corresponding model are depicted in the left column of the first four
    rows. Each component's corresponding Bayesian generalized Lomb-Scargle
    periodogram is depicted in the adjacent right column with the
    \emph{vertical dashed lines} highlighting the orbital period of \name{} b
    ($P=3.44$ days), the stellar rotation period (\prot{} $=44.7$ days), and
    its first harmonic (\prot{}$/2 = 22.3$ days).
    \emph{First row}: the raw RVs. \emph{Second row}: RV activity at \prot{}$/2$
    modeled as separate quasi-periodic GPs for each spectrograph.
    \emph{Third row}: the Keplerian orbital solution for \name{} b.
    \emph{Fourth row}: the RV residuals. \emph{Bottom panel}: the phase-folded
    RV signal from \name{} b.}
  \label{fig:rvs}
\end{figure*}

The raw HARPS-N and HIRES RVs are shown in the top row of \autoref{fig:rvs}
along with their Bayesian generalized Lomb-Scargle periodogram
\citep[\texttt{BGLS};][]{mortier15}. The periodicity induced by \name{} b
is distinctly visible at 3.44 days. A preliminary RV analysis indicated that
following the removal of an optimized Keplerian solution for \name{} b, the
BGLS revealed a strong periodic signal at 22 days, which is seen at
moderately low significance in the BGLS of the raws RVs in \autoref{fig:rvs}.
This periodicity is close to the first harmonic of the stellar rotation period
at \prot{}$/2 \approx 22.3$ days. As such, we interpret this signal as likely
being produced by active regions on the rotating stellar surface.
We note that this feature at \prot{}$/2$ is similar to the first
harmonic of \prot{} observed on the Sun that has been shown to have either a
comparable amount, or at times more power than at \prot{}
\citep{mortier17,milbourne19}. However, we note that
simulated RV time series with injected quasi-periodic magnetic activity signals
have been shown to produce spurious, and sometimes long-lived, periodogram
signals that can masquerade as rotation signatures \citep{nava20}. But given
that the 22-day signal is nearly identical to the first harmonic of the
measured
rotation period, we proceed with treating the 22-day signal as stellar activity
and opt to simultaneously fit the HARPS-N and HIRES RVs with model components
for \name{} b, in the form of a Keplerian orbit, plus a quasi-periodic GP
regression model of stellar activity whose covariance kernel as a function
of time $t$ takes the form

\begin{equation}
  k_{ijs} = a_s^2 \left[ -\frac{(t_i-t_j)^2}{2\lambda^2} - \Gamma^2 \sin^2{\left( \frac{2 \pi |t_i-t_j|}{P_{\text{rot}}} \right)} \right].
\end{equation}

\noindent The quasi-periodic kernel is parameterized by four hyperparameters:
the covariance amplitude $a_s$, where $s$ is the index over the two
spectrographs, the exponential timescale $\lambda$, the coherence
$\Gamma$, and the periodic timescale, which we initialize to \prot{}$/2$ because
of its apparent periodicity in the BGLS of the raw RVs.
Because the temporally-correlated signal that we are modeling with a GP likely
originates from active regions on the rotating stellar surface, and
the fact that activity signals are inherently chromatic, we consider separate
GP activity models for each spectrograph. We also maintain that the 
covariance hyperparameters $\{ \lambda, \Gamma, P_{\text{rot}} \}$ are identical
within each spectrograph's GP activity model. We include an additive
scalar jitter $s_{\text{RV},s}$ for each spectrograph to account for any excess
noise in the activity model and fit for each spectrograph's unique zero-point
offset $\gamma_s$.

Our full RV model therefore consists of 14 free parameters:
$\{ \ln{a_{\text{HARPS-N}}}, \ln{a_{\text{HIRES}}}, \ln{\lambda}, \ln{\Gamma}, \ln{P_{\text{rot}}},$
$\ln{s_{\text{RV,HARPS-N}}}, \ln{s_{\text{RV,HIRES}}}, \gamma_{\text{HARPS-N}}, \gamma_{\text{HIRES}}, P, T_0,$
$\ln{K}, h=\sqrt{e}\cos{\omega}, k=\sqrt{e}\sin{\omega} \}$ where
$K$ is the RV semi-amplitude of \name{} b. The adopted model parameter priors
are included in \autoref{tab:priors}. We fit the RV data with our full
model using the affine invariant ensemble MCMC sampler \texttt{emcee}
\citep{foremanmackey13}, throughout which we use the \texttt{george} package
\citep{ambikasaran14} to evaluate the marginalized likelihood of the GP
activity models. MAP point estimates of the model parameters are derived from
their respective marginalized posterior PDFs and are reported in
\autoref{tab:results}.
  
The second row in \autoref{fig:rvs} depicts the activity component in our RV
model after the MAP Keplerian solution for \name{} b is subtracted from the raw
RVs. The residual periodicity close to \prot{}$/2=22$ days becomes clearly
visible in the BGLS of the RV activity signal. In our GP activity model, we
measure an exponential timescale of
$\lambda=115\pm 20$ days indicating that active regions are relatively stable
over a few rotation cycles. According to detailed investigations of periodogram
signals in simulated RV time series, the persistence of the
maximum RV activity peak at \prot{}$/2$ is consistent with
active region lifetimes on \name{} exceeding \prot{} \citep{nava20}.

In the third row of \autoref{fig:rvs}, the BGLS of the \name{} b signal
is clearly dominated by the 3.44-day periodicity as expected. We measure an RV
semi-amplitude of $K=4.11^{+0.43}_{-0.50}$ \mps{,} which is detected at
$8.2\sigma$ and is clearly visible in the phase-folded RVs in
\autoref{fig:rvs}. The RV residuals, after removing each spectrograph's
mean GP activity model and the MAP Keplerian solution, show no signs of any
probable periodicities and have rms values of 1.90 and 1.65 \mps{} for
the HARPS-N and HIRES RVs respectively. We note that these rms values exceed
the typical RV measurement uncertainties of 1.2 \mps{} and may be indicative
of an incomplete RV model. We reserve an exploration of this prospect until
Sect.~\ref{sect:two}.

\subsection{A global transit \& RV analysis} \label{sect:exofast}
To assess the robustness of the parameters derived from our fiducial modeling
strategy (Sects.~\ref{sect:tessmcmc} and~\ref{sect:rvs}),
here we consider an alternative global model using the
\texttt{EXOFASTv2} exoplanet transit plus RV fitting package
\citep{eastman13,eastman19}.

Here we highlight a few notable differences between our fiducial analysis and
the global model using \texttt{EXOFASTv2}. In our fiducial model of the \tess{}
\texttt{PDCSAP} light curve, we simultaneously fit the data with a GP
detrending model plus a transit model such that the uncertainties in the
recovered planetary parameters are marginalized over our uncertainties in the
detrending model. Conversely, \texttt{EXOFASTv2} takes as input a pre-detrended
light curve to which the transit model is fit. We construct the detrended light
curve to supply to \texttt{EXOFASTv2} using the mean function of the
predictive GP distribution shown in \autoref{fig:tess}. With this method,
the uncertainties in the
planetary parameters of interest are not marginalized over uncertainties in the
detrending model and may consequently be underestimated. Similarly, the RV model
in our fiducial analysis considers temporally-correlated RV activity signals and
models them as a quasi-periodic GP. Conversely, modeling of the prominent
22-day signal in the RVs with \texttt{EXOFAST} requires one to assume a
deterministic functional form for the signal in order to construct a more
complete RV model.
For this purpose, we model the 22-day signal as an eccentric Keplerian
within \texttt{EXOFASTv2}. We adopt broad uniform priors on the signal's $P$ and
$T_0$ and adopt identical priors on its semi-amplitude, $e\cos{\omega}$,
$e\sin{\omega}$ as are used for \name{} b (\autoref{tab:priors}).

The \texttt{EXOFASTv2} model has the important distinction of
evaluating a global model that jointly considers the \tess{} 
photometry along with the HARPS-N and HIRES RVs. By virtue of this, the common
planet parameters between these datasets (i.e. $P$, $T_0$, $e$, $\omega$) will
be self-consistent. In particular, the eccentricity of \name{} b will be jointly
constrained by the transit duration, the RV solution, and the stellar
density, which is constrained by our priors on the stellar mass and radius
(\autoref{tab:priors}). The \texttt{EXOFASTv2} software also explicitly fits for
any excess photometric dilution therefore providing an improved accuracy on the
transit depth and hence on the recovered planetary radius. Within
\texttt{EXOFASTv2}, the dilution is defined as the fractional flux contribution
from neighboring stars (see Section 12 \citealt{eastman19}).

We report the results from our global model in \autoref{tab:results} and compare
the planetary parameters to those derived from our fiducial analysis. All
planetary parameters are consistent between our two analysis strategies at
$<1\sigma$. In particular, in our fiducial and \texttt{EXOFASTv2} analyses, we
measure consistent values for the observables $r_p/R_s=0.0254\pm 0.0009$ and
$0.0257\pm 0.0007$ and $\ln{K}=1.41^{+0.10}_{-0.13}$ and
$1.46^{+0.11}_{-0.13}$.
Given the identical stellar parameter priors in each analysis, this consistency
directly translates into consistent measures of \name{} b's fundamental
planet parameters.

\section{Discussion} \label{sect:discussion}
\subsection{Fundamental planet parameters}
\subsubsection{Orbital separation, mass, and radius}
Our analysis of the \tess{} \texttt{PDCSAP} light curve reveals that \name{}
b has an orbital period of $P=3.444729^{+0.000031}_{-0.000028}$ days and a
planetary radius of $r_p=1.738^{+0.087}_{-0.076}$ \Rearth{.} The corresponding
semimajor axis for \name{} b is $a=0.03845^{+0.00037}_{-0.00040}$ AU where it
receives $53.6^{+5.3}_{-4.7}$ times Earth's insolation.
Assuming uniform heat redistribution and a Bond albedo of zero, \name{} b has
an equilibrium temperature of \teq{}$=754\pm 18$ K.

\begin{figure}
  \centering
  \includegraphics[width=\hsize]{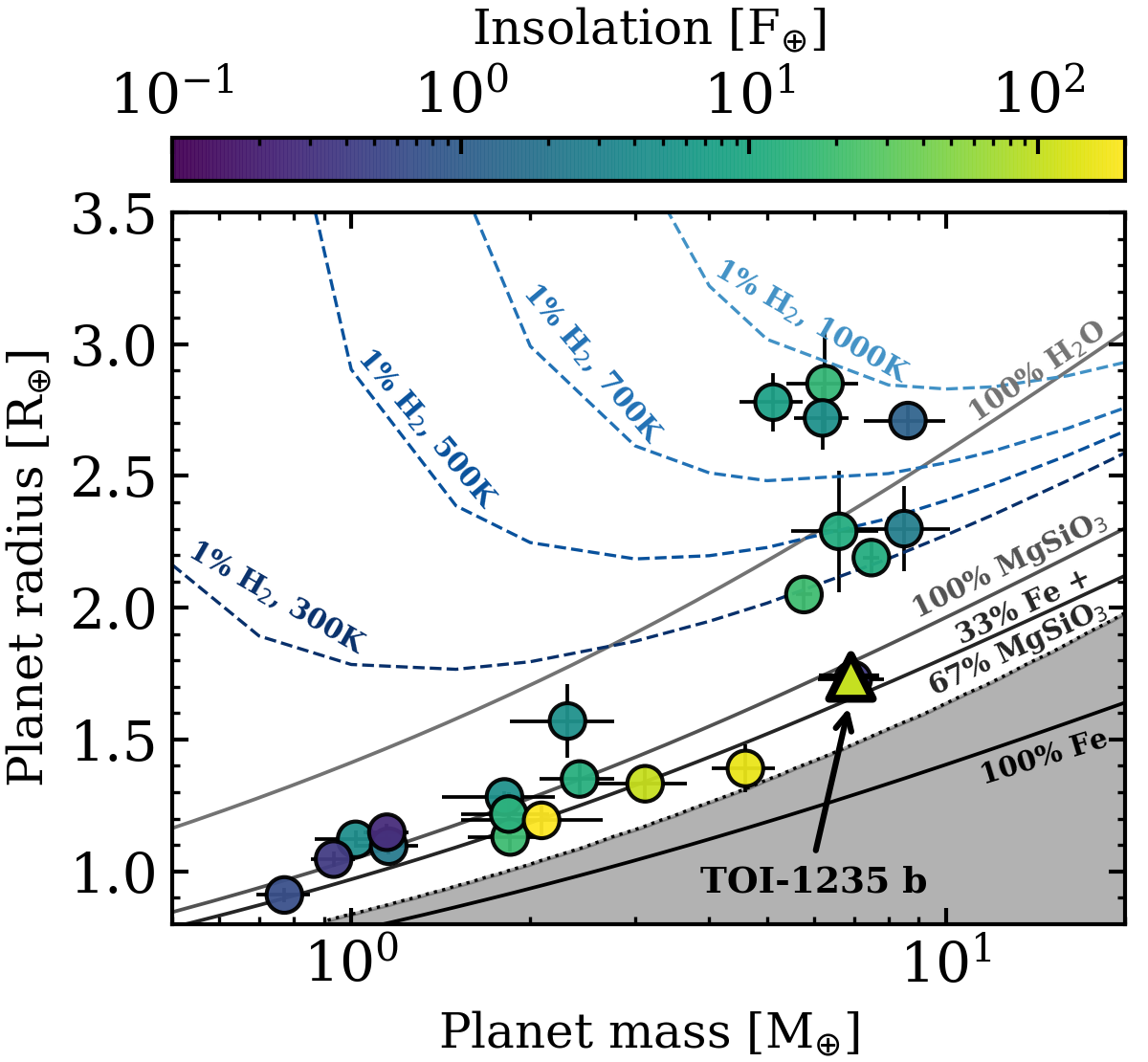}
  \caption{Mass-radius diagram for small planets orbiting M dwarfs including
    \name{} b (\emph{triangle marker}). Errorbars on the \name{} b mass and
    radius are smaller than the marker, which lies directly on top of LHS 1140 b
    in this space. The \emph{solid curves} depict internal
    structure models with mass fractions of 100\% water, 100\% silicate rock,
    33\% iron plus 67\% rock (i.e. Earth-like), and 100\% iron \citep{zeng13}.
    The \emph{dashed curves} depict models of Earth-like cores hosting
    H$_2$ 
    envelopes with 1\% envelope mass fractions at various equilibrium
    temperatures \citep{zeng19}. The \emph{shaded region} bounded by the
    \emph{dotted curve} highlights the forbidden region according to
    models of maximum collisional mantle stripping by giant impacts
    \citep{marcus10}.}
  \label{fig:mr}
\end{figure}

From our RV analysis, we obtain a $8.1\sigma$ planetary mass measurement of
$m_p=6.91^{+0.75}_{-0.85}$ \Mearth{.} Taken together, the mass and radius of
\name{} b give a bulk density of $\rho_p=7.4^{+1.5}_{-1.3}$ g cm$^{-3}$. In
\autoref{fig:mr} we add \name{} b to the mass-radius
diagram of small M dwarf planets with $\geq 3\sigma$ mass measurements.
Comparing \name{} b's mass and radius to internal
structure models of two-layer, fully-differentiated planet interiors
\citep{zeng13} reveals that the bulk composition of \name{} b is consistent
with an Earth-like composition of 33\% iron plus 67\%
silicate rock by mass.

Intriguingly, the mass and radius of \name{} b are nearly
identical to those of LHS 1140 b despite LHS 1140 b having a wider 25-day orbit 
around a mid-M dwarf, thus making it much more temperate than \name{} b
\citep[\teq{}$=230$ K;][]{dittmann17a,ment19}. Both planets are situated within
the radius valley around low mass stars \citepalias{cloutier20}
and have masses that
appear to represent the upper limit of terrestrial planet masses in a planetary
mass regime where rocky Earth-like planets are inherently rare (i.e. 5-10
\Mearth{,} \autoref{fig:mr}). These planets offer unique opportunities to study
nature's largest terrestrial planets whose tectonic and outgassing
processes may differ significantly from those on Earth-sized terrestrial
planets \citep{valencia07b}.

With the planetary
mass measurement presented herein, \name{} adds to the growing list
of small planets transiting M dwarfs with precise RV masses
(GJ 3470; \citealt{bonfils12}, GJ 1214; \citealt{charbonneau09},
GJ 1132; \citealt{bonfils18}, K2-3; \citealt{damasso18},
K2-18; \citealt{cloutier19a}, LHS 1140; \citealt{ment19})
that has been rapidly expanding since the launch of \tess{}
(GJ 357; \citealt{luque19}, GJ 1252; \citealt{shporer20},
L 98-59; \citealt{cloutier19c}, L 168-9; \citealt{astudillodefru20},
LTT 3780; \citealt{cloutier20b,nowak20}). Notably, \name{} b also directly
contributes to the completion of the \tess{} level one science
requirement of obtaining precise masses for fifty planets smaller than
four Earth radii.

\subsubsection{Iron and envelope mass fractions}
We wish to place self-consistent limits on the iron mass fraction
$X_{\text{Fe}}$ and envelope mass fraction $X_{\text{env}}$ of \name{} b.
Here the iron mass fraction is defined as the ratio of the total mass of the
core and mantle that is composed of iron, with the remainder in magnesium
silicate. 
The envelope mass fraction is then defined as the fraction of the planet's total
mass that is in its gaseous envelope.
However, it is important to note that these values
are degenerate such that we cannot derive a unique solution given only the
planet's mass and radius. For example, the bulk composition of
\name{} b is consistent with being Earth-like, thus suggesting a small envelope
mass fraction\footnote{The Earth has an envelope mass fraction of
  $<10^{-6}$.}, but one could also imagine a more exotic scenario that is
consistent with the planet's mass and radius of a  planetary core
with $X_{\text{Fe}}=1$,
surrounded by an extended H/He envelope. In the simplest case, we assume
that magnesium silicate and iron are the only major constituents of
\name{} b's bulk composition such that $X_{\text{env}}=0$. Under this
assumption, we derive $X_{\text{Fe}}$ by Monte-Carlo sampling the uncorrelated
marginalized posterior PDFs of $m_p$ and $r_p$ and use the analytical
rock/iron mass-radius relation from \cite{fortney07} to recover
$X_{\text{Fe}}$. We find that \name{} b has an iron
mass fraction of $X_{\text{Fe}}=20^{+15}_{-12}$\% that is $<46$\% at 90\%
confidence. 

\begin{figure}
  \centering
  \includegraphics[width=\hsize]{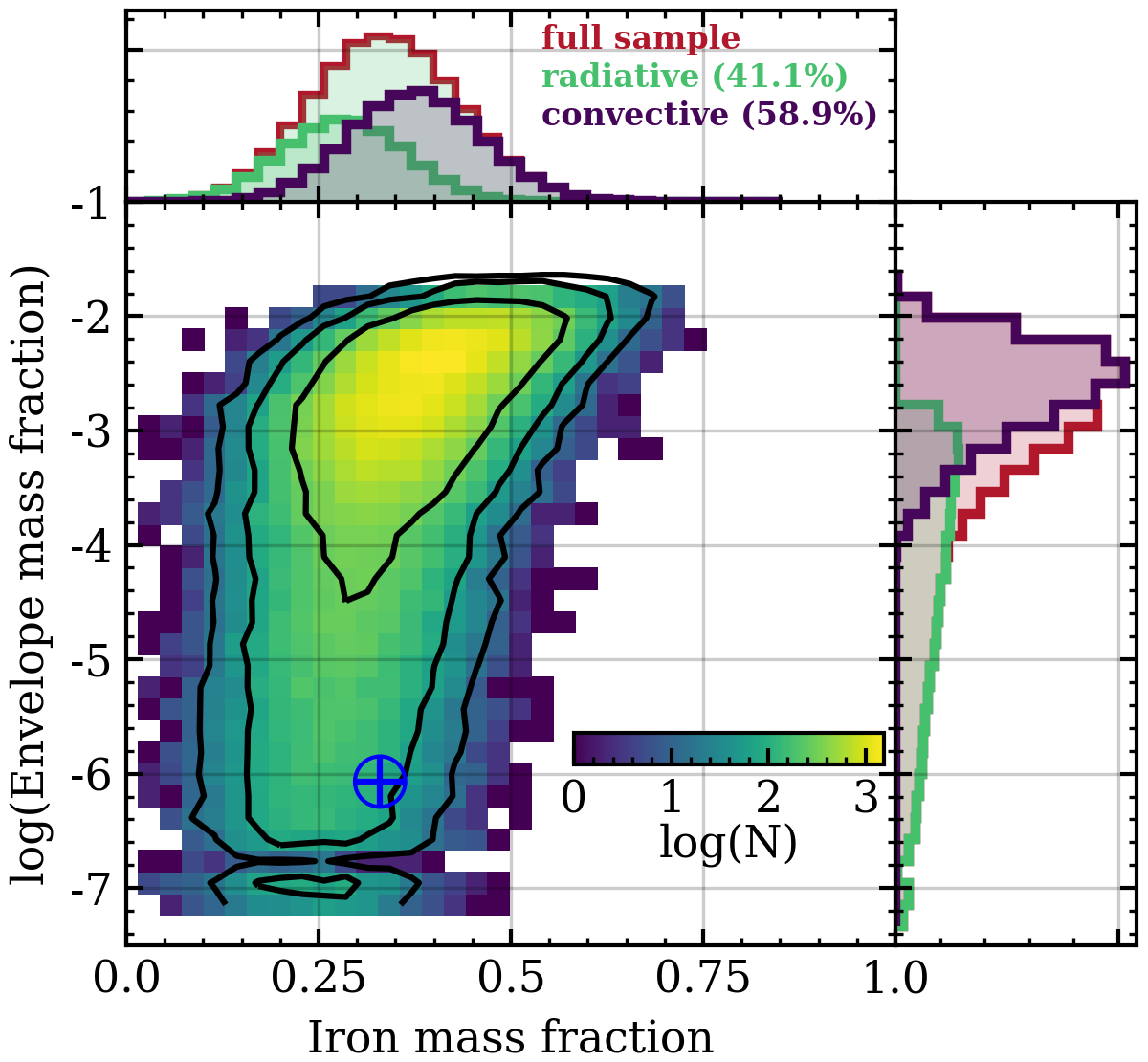}
  \caption{Joint distribution of \name{} b's iron mass and envelope mass
    fractions
    to be consistent with its measured mass and radius. The colormap represents
    the number of successful planet models given $X_{\text{env}}$ and
    $X_{\text{Fe}}$ while the contours highlight the 68, 95, and 99
    percentiles. The top and right 1D histograms depict the marginalized
    distributions of $X_{\text{Fe}}$ and $\log_{10}{(X_{\text{env}})}$
    respectively for the full sample (\emph{red}), plus the subset of
    realizations with a radiative atmospheres (\emph{green}) and
    convective atmospheres (\emph{blue}). The \emph{blue cross} highlights the
    Earth.}
  \label{fig:Xenv}
\end{figure}

To infer the distribution of envelope mass fractions that are consistent with
the data, we first impose a physically-motivated prior on $X_{\text{Fe}}$
of $\mathcal{N}(0.33,0.10)$. The relatively narrow width of this Gaussian
prior is qualitatively supported by observations of nearby Sun-like stellar
metallicities that show that the abundance ratios of Mg/Fe, Si/Fe, and Mg/Si
at similar ages and metallicities vary by less than $10$\%. This indicates
a low level of compositional diversity in the refractory building blocks of
planets \citep{bedell18}. The width of our $X_{\text{Fe}}$ Gaussian prior is
chosen in an ad hoc way to approximately reflect this level of chemical
diversity. The
homogeneity of refractory chemical abundances among Sun-like stars, coupled
with their similar condensation temperatures \citep{lodders03},
suggests a narrow range in
iron mass fractions among close-in terrestrial planets. This assertion is
supported by the locus of terrestrial planets with $r_p\lesssim 1.8$ \Rearth{}
that are consistent with an Earth-like bulk composition (\autoref{fig:mr}).
This concept of similar $X_{\text{Fe}}$ values is particularly compelling for
the most massive terrestrial planets (e.g. \name{} b) for which
a significant increase in $X_{\text{Fe}}$ by collisional mantle stripping
is energetically infeasible due to the large binding energies of such planets
\citep{marcus10}.

To proceed with deriving the distribution of \name{} b envelope mass fractions
assuming an Earth-like core, we extend the solid two-layer interior structure
model to include a H/He envelope with a mean molecular weight equal to
that of a solar metallicity gas ($\mu=2.35$). Our adopted
planetary model is commonly used for sub-Neptune-sized
planets \citep[e.g.][]{rafikov06,lee15,ginzburg16,owen17,gupta19}.
This model features a solid core surrounded by a H/He gaseous
envelope that, depending on the planetary parameters, is either fully radiative
throughout or may be convective in the deep interior up to the height of the
radiative-convective boundary (RCB), above which the atmosphere becomes
radiative and isothermal with temperature \teq{.} The latter scenario
represents the general case whereas the former is only invoked when the 
planetary parameters result in a height of the RCB that is less than the
atmospheric pressure scale height at \teq{.}
To first order, the height of the RCB above the planetary surface
$r_{\text{RCB}}$, and hence $X_{\text{env}}$, are determined by
$\{ T_{\text{eq}}, m_p, r_p, X_{\text{Fe}} \}$.
Each of \teq{,} $m_p$, and $r_p$ are directly constrained by our
data if we assume a Bond albedo to infer \teq{.}
We derive $r_{\text{RCB}}$ and $X_{\text{env}}$
by Monte-Carlo sampling $X_{\text{Fe}}$ from its prior,
along with the zero-albedo
\teq{,} $m_p$, and $r_p$ from their respective marginalized posterior PDFs.
We then rescale each \teq{} draw by $(1-A_B)^{1/4}$ where $A_B$ is the Bond
albedo. Super-Earth bond albedos have poor empirical constraints so we opt to
condition a broad uniform prior on $A_B$ of $\mathcal{U}(0.1,0.8)$ based on
the solar system planets. Lastly,
although we expect $r_{\text{RCB}}$ to shrink over
time as the H/He envelope cools and contracts, this effect on
$X_{\text{env}}$ is known to be a weak function of planet age \citep{owen17}
such that we fix the age of \name{} to 5 Gyrs in our calculations.


We use a customized version of the \texttt{EvapMass}
software \citep{owen20} to self-consistently solve
for $r_{\text{RCB}}$ and $X_{\text{env}}$ given samples
of $\{ T_{\text{eq}}, m_p, r_p, X_{\text{Fe}} \}$. We attempt to sample these
parameters in 
$10^5$ realizations although not all parameter combinations are physically
capable of producing a self-consistent solution.
In practice, our Monte-Carlo sampling results in 94,131  successful planetary
model realizations (i.e. 94.1\% success rate). The resulting distributions of
$X_{\text{Fe}}$ and $X_{\text{env}}$ that are consistent with our measurements of
\name{} b are shown in \autoref{fig:Xenv}. 
We find that 41.1\% of successful planet model realizations have fully radiative
atmospheres
with the remaining 58.9\% being convective in the lower atmosphere. These models
produce largely disparate results with radiative atmospheres being favored
for increasingly smaller $X_{\text{Fe}}$ and always having
$X_{\text{env}} \lesssim 10^{-3}$. Conversely, atmospheres with a deep convective
region are more extended thus requiring a more compressed core (i.e. large
$X_{\text{Fe}}$) and larger $X_{\text{env}}$.
Overall we see the positive correlation between $X_{\text{Fe}}$
and $X_{\text{env}}$ because at a fixed $m_p$, the core radius must shrink with
increasing $X_{\text{Fe}}$, which requires the envelope to become extended to
match the observed radius. Extending the envelope increases
the limits of integration over the atmospheric density profile from the
planetary surface to the top of the atmosphere, consequently increasing the
envelope mass. With our models, we find that \name{} b has a maximum envelope
mass fraction of 2.3\%.
Marginalizing over all other model parameters, and both atmospheric equations
of state, we find that
$X_{\text{env}}$ must be $<0.5$\% at 90\% confidence.

\subsection{Implications for the origin of the radius valley around mid-M
  dwarfs}
Observational studies of the occurrence rate of close-in planets around Sun-like
stars have revealed a bimodality in the distribution of planetary radii known
as the radius valley \citep[e.g.][]{fulton17,fulton18,mayo18}. This dearth of
planets between 1.7-2.0 \Rearth{} around Sun-like stars
likely marks the transition between rocky
planets and larger planets that host extended gaseous envelopes. Physical
models of the emergence of the radius valley from thermally-driven atmospheric
mass loss (i.e. photoevaporation or core-powered mass loss),
and from terrestrial planet formation in a gas poor environment, make
distinct predictions regarding the slope of the radius valley in period-radius
space. The slope of the radius valley around Sun-like stars with \teff{} $>4700$
K was measured by \cite{martinez19} using the stellar sample from
\cite{fulton17}. The recovered slope was shown to be consistent with model
predictions from thermally-driven atmospheric mass loss 
\citep[$r_{p,\text{valley}}\propto P^{-0.15}$;][]{lopez18}.
On the other hand, the slope around lower mass dwarfs with \teff{} $<4700$ K
(i.e. mid-K to mid-M dwarfs) was measured by \citetalias{cloutier20} and was
shown to have a flipped sign that instead was consistent with predictions from
gas-poor formation \citep[$r_{p,\text{valley}}\propto P^{0.11}$;][]{lopez18}.
One interpretation of this is that the dominant mechanism for
sculpting the radius valley is stellar mass dependent and that thermally-driven
mass loss becomes less efficient towards mid-to-late M dwarfs where a new
formation pathway of terrestrial planets in a gas-poor environment emerges
\citepalias{cloutier20}.
The stellar mass at which this proposed transition occurs is not
well resolved by occurrence rate measurements, but it may be addressed by the
detailed characterization of individual planets that span the model predictions
in period-radius space (e.g. \name{} b).

\begin{figure}
  \centering
  \includegraphics[width=\hsize]{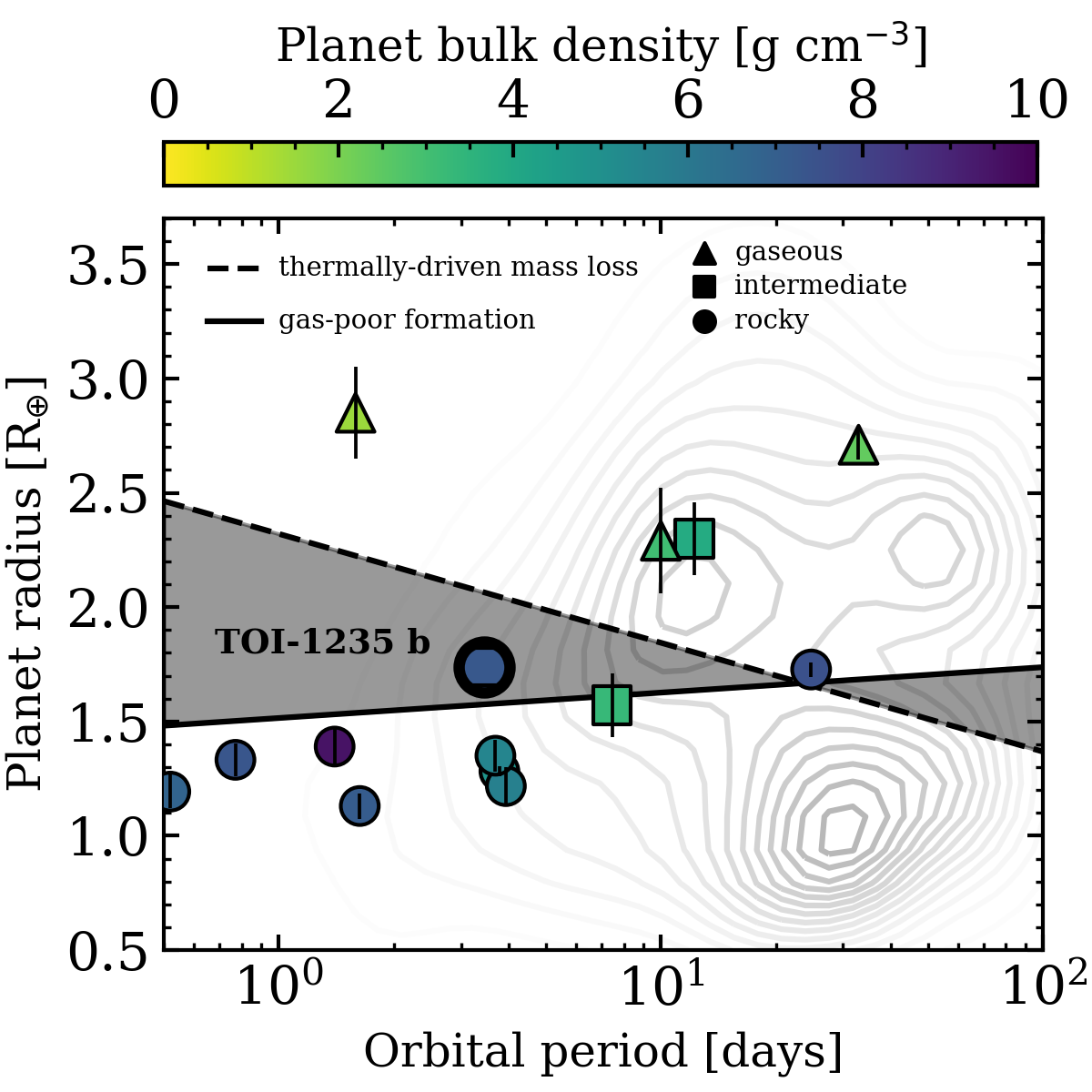}
  \caption{Period-radius diagram for small planets orbiting M dwarfs with
    precise RV masses including \name{} b (\emph{bold circle}). The
    \emph{dashed and solid lines} depict the locations of the radius
    valley around low mass stars from model predictions of thermally-driven
    atmospheric mass loss and from gas-poor terrestrial planet formation
    respectively. The shaded regions highlight where the
    predictions of planetary bulk compositions are discrepant between the two
    models. Contours represent the
    planetary occurrence rates around low mass stars \citepalias{cloutier20}.
    Planet marker shapes depict the planet's compositional disposition
    as either rocky (\emph{circles}), gaseous (\emph{triangles}), or
    intermediate (\emph{squares}). Marker colors indicate the planet's bulk
    density.}
  \label{fig:radval}
\end{figure}

Differences in the slopes of the radius valley around Sun-like and lower mass
stars naturally carve out a subset of the
period-radius space in which the models make opposing predictions for the
bulk compositions of planets. This subspace around low mass stars cooler than
4700 K was quantified by \citetalias{cloutier20} and is highlighted in
\autoref{fig:radval}. At periods less than 23.5 days, planets within the
highlighted subspace are expected to be rocky according to models of
thermally-driven hydrodynamic escape. Conversely, gas-poor formation models
predict that those planets should instead be non-rocky with envelope mass
fractions of at least a few percent depending on their composition.
\name{} b falls within this region of
interest and therefore provides direct constraints on the efficiency of the
competing physical processes on close-in planets around early M dwarfs.

Our transit and RV analyses revealed that \name{} b is a predominantly rocky
planet with an iron mass fraction of $20^{+15}_{-12}$\% 
and an envelope mass fraction that is $<0.5$\% at 90\% confidence.
Given its period and radius, this finding is consistent with models of
thermally-driven mass loss but is inconsistent with the gas-poor formation
scenario. Indeed, based on the photoevaporation-driven hydrodynamic escape
simulations by \cite{lopez13}, the mass of \name{} b place its insolation
flux ($F=53.6^{+5.3}_{-4.7}$ F$_{\oplus}$) right at the
threshold insolation required for the planet to lose its gaseous envelope:
$F_{\text{threshold}}=52\pm 14$ F$_{\oplus}$.\footnote{Assuming a fixed mass loss
  efficiency of 10\% \citep{lopez12}.}
These results suggest that thermally-driven mass loss
continues to be an efficient
process for sculpting the radius valley around early M dwarfs like \name{.}
\citetalias{cloutier20} suggested that although thermally-driven mass loss
seems to be prevalent around Sun-like stars,
evolution in the structure of the radius valley with stellar mass suggests that
this prevelance weakens with decreasing stellar mass and that
gas-poor formation may emerge as the dominant mechanism for sculpting the
radius valley around early-to-mid M dwarfs. Although the stellar mass at which
this proposed transition occurs has yet to be resolved, the rocky nature of
\name{} b further suggests that the stellar mass at which this transition occurs
is likely less than that of \name{} ($0.640\pm 0.016$ \Msun{)}.

As an aside, we note that distinguishing between photoevaporation and
core-powered mass loss cannot be achieved with the data presented herein.
Fortunately, the distinction can be addressed at the
planet population level by investigating the radius valley's dependence with
time and with stellar mass \citep{gupta20}.

\subsection{Testing the prospect of a second planet around \name{}} \label{sect:two}
Recall that after removing the \name{} b signal from our RV time
series, a strong residual periodicity emerges at about 22 days
(second row in \autoref{fig:rvs}). We initially interpreted this signal as
being likely
related to rotationally-induced stellar activity because of its proximity
to the first harmonic of the probable stellar rotation period inferred from
ground-based photometric monitoring (\prot{}$=44.7\pm 4.5$ days,
\autoref{fig:prot}). Although the measurement of \prot{} makes the 22-day RV
signal suggestive of being related to stellar activity, here we conduct a
suite of tests that instead favor a planetary origin.

The treatment of the 22-day RV signal as either a quasi-periodic GP in our
fiducial model or as an eccentric Keplerian in our \texttt{EXOFASTv2} global
model (see Sects.~\ref{sect:rvs} and~\ref{sect:exofast}), gives an activity
semi-amplitude of $\approx 5$ \mps{.} This value appears to be at odds with
reasonable predictions of the RV signal based on the star's long-term
photometric variability from ground-based monitoring (Sect.~\ref{sect:prot}).
Using the $FF'$ model
to predict the activity-induced RV variations from photometric variability 
\citep{aigrain12}, we would expect the semi-amplitude of the \name{} RV activity
signal to be at the level of 1-2 \mps{} instead of the observed value of 5
\mps{} under the single-planet model. However, it is important to note that
photometry is not a perfect predictor of RV variations because i) stellar
activity undergoes cycles and there is no guarantee that the level of activity
is constant between the epochs of photometric monitoring and the RV
observations, ii) photometry is not sensitive to all spot distributions
\citep{aigrain12} and iii) bright chromospheric plages can produce RV
variations with amplitudes similar to those induced by spots of the same size,
but with potentially ten times less flux variations \citep{dumusque14}.
Therefore, the discrepancy between the observed RV activity variations and the
$FF'$ model predictions is merely suggestive that
our RV activity models are over-predicting the amplitude of the RV activity
signal, which would then require an additional RV component to model the excess
signal in the RV residuals.

Rotationally-induced RV signals from active regions arise from the temperature
difference between the active regions and the surrounding stellar surface. As
such, the active region contrast has an inherent wavelength dependence
that increases towards shorter wavelengths such that RV activity signals should
be larger at bluer wavelengths \citep{reiners10}. We elected to investigate
the chromatic dependence of the 22-day RV signal by considering sets of `blue'
and `red' RVs from HARPS-N. We re-derived the HARPS-N RVs using the same
methodology as in our fiducial analysis but focused separately on the
spectral orders 0-45 (388-550 nm) and 46-68 (550-689 nm)
to derive sets of blue and red RVs respectively. Each range of orders was
selected to achieve a comparable median RV measurement uncertainty in each
time series of 1.98 \mps{} and 2.04 \mps{.} We then investigated the
chromatic dependence of the probability of the 3.44-day and 22-day periodicities
in the BGLS. While the 22-day signal was marginally more probable in the red RVs,
we found that 
the planetary signal varied by many more orders-of-magnitude than the 22-day signal.
This behavior is unexpected for a planetary signal that is known to be
achromatic. We therefore concluded that this chromatic analysis of our dataset is
unreliable and we make no claims regarding the physical origin of the 22-day signal
based on its chromatic dependence.

To explicitly test the idea that an additional RV component is required to
completely model the data,
we considered a two-planet RV model with components for \name{} b, a
second Keplerian `c' at 22 days, plus quasi-periodic GP activity models
for each spectrograph
with an imposed prior on its periodic timescale equal to that of \prot{:}
$\mathcal{N}(44.7,4.5)$ days. 
We sampled the two-planet model parameter posteriors using an identical method
to what was used in our fiducial analysis of the one-planet RV model
(Sect.~\ref{sect:rvs}). We adopted narrow uniform priors on
$P_c$ of $\mathcal{U}(17,27)$ days and on
$T_{0,c}$ of $\mathcal{U}(1821.5,1848.5)$ BJD - 2,457,000.
The resulting Keplerian model parameters on the hypothetical planet `c' are
reported in \autoref{tab:c}. We find that the hypothetical planet would have
a period of $P_c=21.8^{+0.9}_{-0.8}$ days and an RV semi-amplitude of
$K_c=4.2^{+1.2}_{-1.7}$, which implies a minimum mass
of $m_{p,c}\sin{i}=13.0^{+3.8}_{-5.3}$ \Mearth{.}

\begin{deluxetable}{lc}
\tabletypesize{\footnotesize}
\tablecaption{Point estimates of the hypthetical planet `c' Keplerian model parameters\label{tab:c}}
\tablewidth{0pt}
\tablehead{Parameter & Model Values}
\startdata
Orbital period, $P_c$ [days] & $21.8^{+0.9}_{-0.8}$ \\
Time of mid-transit, $T_{0,c}$ [BJD - 2,457,000] & $1835.3^{+2.2}_{-2.1}$ \\
Log RV semi-amplitude, $\ln{K_c/m/s}$ & $1.4^{+0.3}_{-0.5}$ \\
$\sqrt{e_c}\cos{\omega_c}$ & $0.08^{+0.3}_{-0.4}$  \\
$\sqrt{e_c}\sin{\omega_c}$ & $0.12^{+0.37}_{-0.47}$ \\
\multicolumn{2}{c}{\emph{Derived parameters}} \\
RV semi-amplitude, $K_c$ [\mps{]} & $4.2^{+1.2}_{-1.7}$ \\
Minimum planet mass, $m_{p,c}\sin{i}$ [\Mearth{]} & $13.0^{+3.8}_{-5.3}$ \\
Semimajor axis, $a_c$ [AU] & $0.1319^{+0.0046}_{-0.0043}$ \\
Insolation, $F_c$ [F$_{\oplus}$] & $4.6^{+0.6}_{-0.5}$ \\
Equilibrium temperature, $T_{\text{eq},c}$ [K] & \\
\hspace{2pt} Bond albedo = 0.0 & $407\pm 12$ \\
\hspace{2pt} Bond albedo = 0.3 & $373\pm 11$ \\
\enddata
\tablecomments{Note that we do not conclude that the hypothetical planet `c' presented in this table is a bona-fide planet.}
\end{deluxetable}

We now have one and two-planet RV models of the HARPS-N plus HIRES RVs that
both include a GP activity component whose periodic time scales are constrained
to be close to \prot{}$/2$ and \prot{} respectively. Therefore,
we can use our models to
conduct a model comparison to assess the favorability of one model over the
other. We used the marginalized
posterior PDFs from each model's MCMC results to estimate their Bayesian model
evidences $\mathcal{Z}$ using the estimator from \cite{perrakis14}.
We estimate model evidences of $\ln{\mathcal{Z}_1}=-110.0$
and $\ln{\mathcal{Z}_2}=-91.0$, which gives a model evidence ratio of
$\mathcal{Z}_2/\mathcal{Z}_1=10^8$. This result strongly favors the
two-planet model although we caution that Bayesian model evidences are
notoriously difficult to accurately calculate and their interpretation is
dependent on the assumed model parameter priors \citep{nelson18}.
Alternatively, we also compute the Bayesian
information criterion (BIC) and the Akaike information criterion (AIC) to
perform model comparisons that are independent of the model priors. We measure
BIC$_1=225.0$ and BIC$_2=205.2$ such that the two-planet model is again strongly
favored since
$\Delta \text{BIC}_{12} \equiv \text{BIC}_1 - \text{BIC}_2 =19.8 > 10$.
This is further supported by AIC$_1=202.1$ and AIC$_2=174.1$ whereby the
two-planet model remains strongly favored as
$\Delta \text{AIC}_{12} \equiv \text{AIC}_1 - \text{AIC}_2 = 28.0 > 10$.

Encouraged by the prospect of a second planet orbiting \name{,} we
used its measured orbital period $P_c=21.8^{+0.9}_{-0.8}$ days and its time of
inferior conjunction $T_{0,c}=2458835.3^{+2.2}_{-2.1}$ BJD to search for
transit-like events in the \tess{} \texttt{PDCSAP} and archival
MEarth-North light curves. With \tess{} we conducted the
search for periodic transit-like signals
close to $P_c$ using the implementation of the
Box Least Squares algorithm \citep[BLS;][]{kovacs02} in \cite{cloutier19b}.
We conducted a complementary BLS search on the full MEarth-North light curve
following the methods outlined in \cite{ment19}.
We do not find any significant transit-like signals other than those associated
with \name{} b. Therefore, if the 22-day signal is truly a planet,
then it is unlikely to be transiting. This result is perhaps unsurprising given
that if the hypothetical planet `c' is
coplanar with \name{} b at $88.1^{\circ}$, then `c' would not have a
transiting configuration at its separation of $a_c/R_s = 45.1^{+2.0}_{-1.9}$.

We emphasize that while the aforementioned lines of evidence are suggestive of
a second, non-transiting planet around \name{,} the data presented herein are
not sufficient to firmly distinguish between planetary and stellar activity
origins of the 22-day RV signal. On-going spectroscopic monitoring of \name{}
over many rotation cycles may help to solve this ambiguity by testing for
temporal correlations of the signal's amplitude over the star's evolving
magnetic activity cycle. A more secure detection of the stellar rotation
period from continued photometric monitoring would also be beneficial.

\subsection{An independent analysis of the \name{} system}
Following the announcement of the TOI-1235.01 level one planet candidate
in October 2019, multiple PRV instrument teams began pursing
its mass characterization through TFOP. This study has
presented the subset of those efforts from HARPS-N and HIRES but
we acknowledge that another collaboration has also submitted a paper presenting
their own RV time series and analysis \citep{bluhm20}.
Although the submissions of these
complementary studies were coordinated between the two groups, their respective
data, analyses, and writeups, were intentionally conducted independently.

\section{Summary} \label{sect:summary}
We have presented the discovery and confirmation of \name{} b, a transiting
super-Earth around a bright early M dwarf from the \tess{} mission. The planet
was confirmed through intensive follow-up observations including a set of
precise RV measurements from HARPS-N and HIRES. The main findings of our study
are summarized below:

\begin{itemize}
\item \name{} is a bright ($V$=11.495, $K_s$=7.893) early M dwarf at 39.6 pc
  with mass and
  radius of $0.640\pm 0.016$ \Msun{} and $0.630\pm 0.015$ \Rsun{.} Archival
  MEarth-North photometry reveals a probable rotation period of $44.7\pm 4.5$
  days.
\item The transiting planet \name{} b has an orbital period of 3.44 days with
  a mass and radius of $6.91^{+0.75}_{-0.85}$ \Mearth{} and
  $1.738^{+0.087}_{-0.076}$ \Rearth{.} \name{} b directly contributes to the
  completion of the \tess{} level one science requirement to deliver masses
  for fifty planets with radii $<4$ \Rearth{.}
\item Planetary structure models reveal that the \name{} b mass and radius are
  consistent with an iron mass fraction of $20^{+15}_{-12}$\% and a H/He
  envelope mass fraction of $<0.5$\% at 90\% confidence, therefore making the
  planet consistent with an Earth-like bulk composition.
\item The period and radius of \name{} b place it between competing model
  predictions of the location of the rocky/non-rocky planet transition.
  The rocky composition of \name{} b makes it consistent with thermally-driven
  atmospheric mass loss scenarios but inconsistent with gas-poor formation
  models suggesting that the former physical process is still
  efficient at sculpting the radius valley around early M dwarfs.
\item We also see a periodic signal in the RV measurements at 22-days,
  close to the first harmonic of the star's probable rotation period.
  While this is suggestive of the signal's origin being related to stellar
  activity, estimates of the RV activity signal's amplitude from photometry
  and the comparison of one and two-planet RV models, suggest that the signals'
  origin may instead be planetary. However, we are unable to definitely
  distinguish between activity and a second planet with the data presented
  herein. 
\end{itemize}

\acknowledgements
RC is supported by a grant from the National Aeronautics and Space
Administration in support of the TESS science mission.
We thank Andrew Vanderburg for enlightening discussions regarding the \tess{}
light curves. 

MPi gratefully acknowledges the support from the European Union Seventh Framework Programme (FP7/2007-2013) under Grant Agreement No. 313014 (ETAEARTH).

JMAM gratefully acknowledges support from the National Science Foundation Graduate Research Fellowship Program under Grant No. DGE-1842400. JMAM also thanks the LSSTC Data Science Fellowship Program, which is funded by LSSTC, NSF Cybertraining Grant \#1829740, the Brinson Foundation, and the Moore Foundation; his participation in the program has benefited this work.

This work has been partially supported by the National Aeronautics and Space
Administration under grant No.  NNX17AB59G issued through the Exoplanets
Research Program.

We acknowledge the use of public TESS Alert data from the pipelines at the
TESS Science Office and at the TESS Science Processing Operations Center.

The MEarth Team gratefully acknowledges funding from the David and Lucile
Packard Fellowship for Science and Engineering (awarded to D.C.). This material
is based upon work supported by the National Science Foundation under grants
AST-0807690, AST-1109468, AST-1004488 (Alan T. Waterman Award), and AST-1616624.
This work is made possible by a grant from the John Templeton Foundation. The
opinions expressed in this publication are those of the authors and do not
necessarily reflect the views of the John Templeton Foundation. This material
is based upon work supported by the National Aeronautics and Space
Administration under Grant No. 80NSSC18K0476 issued through the XRP Program.

This publication makes use of data products from the Two Micron All Sky Survey,
which is a joint project of the University of Massachusetts and the Infrared
Processing and Analysis Center/California Institute of Technology, funded by
the National Aeronautics and Space Administration and the National Science
Foundation.

This work has made use of data from the European Space Agency (ESA) mission
\gaia{} (\url{https://www.cosmos.esa.int/gaia}), processed by the \gaia{} Data
Processing and Analysis Consortium (DPAC,
\url{https://www.cosmos.esa.int/web/gaia/dpac/consortium}). Funding for the
DPAC has been provided by national institutions, in particular the institutions
participating in the \gaia{} Multilateral Agreement.

This work makes use of observations from the LCOGT network.

Based on observations made with the Italian {\it Telescopio Nazionale
Galileo} (TNG) operated by the {\it Fundaci\'on Galileo Galilei} (FGG) of the
{\it Istituto Nazionale di Astrofisica} (INAF) at the
{\it  Observatorio del Roque de los Muchachos} (La Palma, Canary Islands, Spain).

The HARPS-N project has been funded by the Prodex Program of the Swiss Space Office (SSO), the Harvard University Origins of Life Initiative (HUOLI), the Scottish Universities Physics Alliance (SUPA), the University of Geneva, the Smithsonian Astrophysical Observatory (SAO), and the Italian National Astrophysical Institute (INAF), the University of St Andrews, Queen’s University Belfast, and the University of Edinburgh.

Some of the data presented herein were obtained at the W. M. Keck Observatory, which is operated as a scientific partnership among the California Institute of Technology, the University of California, and NASA. The Observatory was made possible by the generous financial support of the W.M. Keck Foundation. The authors wish to recognize and acknowledge the very significant cultural role and reverence that the summit of Maunakea has always had within the indigenous Hawaiian community. We are most fortunate to have the opportunity to conduct observations from this mountain.

Resources supporting this work were provided by the NASA High-End Computing
(HEC) Program through the NASA Advanced Supercomputing (NAS) Division at Ames
Research Center for the production of the SPOC data products.

Based on observations obtained at the international Gemini Observatory under the program GN-2019B-LP-101, a program of NSF’s OIR Lab, which is managed by the Association of Universities for Research in Astronomy (AURA) under a cooperative agreement with the National Science Foundation on behalf of the Gemini Observatory partnership: the National Science Foundation (United States), National Research Council (Canada), Agencia Nacional de Investigaci\'{o}n y Desarrollo (Chile), Ministerio de Ciencia, Tecnolog\'{i}a e Innovaci\'{o}n (Argentina), Minist\'{e}rio da Ci\^{e}ncia, Tecnologia, Inova\c{c}\~{o}es e Comunica\c{c}\~{o}es (Brazil), and Korea Astronomy and Space Science Institute (Republic of Korea). Some of the observations in the paper made use of the High-Resolution Imaging instrument `Alopeke.

\facilities{TESS, MEarth-North, TRES, LCOGT, Gemini/NIRI, TNG/HARPS-N,
  Keck/HIRES.}

\software{\texttt{AstroImageJ} \citep{collins17},
  \texttt{astropy} \citep{astropyi,astropyii},
  \texttt{BANZAI} \citep{mccully18},
  \texttt{batman} \citep{kreidberg15},
  \texttt{BGLS} \citep{mortier15},
  \texttt{celerite} \citep{foremanmackey17},
  \texttt{emcee} \citep{foremanmackey13},
  \texttt{EvapMass} \citep{owen20},
  \texttt{EXOFAST} \citep{eastman13},
  \texttt{EXOFASTv2} \citep{eastman19},
  \texttt{exoplanet} \citep{foremanmackey19},
  \texttt{PyMC3} \citep{salvatier16},
  \texttt{scipy} \citep{scipy},
  \texttt{SpecMatch-Emp} \citep{yee17},
  \texttt{STARRY} \citep{luger19},
  \texttt{Tapir} \citep{jensen13},
  \texttt{TERRA} \citep{anglada12},
  \texttt{triceratops} \citep{giacalone20},
  \texttt{vespa} \citep{morton12}.}

\bibliographystyle{apj}
\bibliography{refs}

\startlongtable
\begin{deluxetable*}{lcc}
\tabletypesize{\footnotesize}
\tablecaption{Point estimates of the \name{} model parameters\label{tab:results}}
\tablewidth{0pt}
\tablehead{Parameter & Fiducial Model Values\tablenotemark{a} & \texttt{EXOFASTv2} Model Values\tablenotemark{b}}
\startdata
\multicolumn{3}{c}{\emph{TESS light curve parameters}} \\
Baseline flux, $f_0$ & $1.000024\pm 0.000010$ & $1.000035\pm 0.000018$  \\
$\ln{\omega_0}$ & $1.45\pm 0.17$ & - \\
$\ln{S_0 \omega_0^4}$ & $-0.16^{+0.52}_{-0.57}$ & - \\
$\ln{s_{\text{TESS}}^2}$ & $0.064\pm 0.006$ & - \\
\tess{} limb darkening coefficient, $u_1$ & $0.47^{+0.32}_{-0.24}$ & $0.40^{+0.34}_{-0.26}$ \\
\tess{} limb darkening coefficient, $u_2$ & $0.20^{+0.38}_{-0.35}$ & $0.23^{+0.37}_{-0.38}$ \\
Dilution & - & $0.09^{+0.21}_{-0.34}$ \\
\smallskip \\
\multicolumn{3}{c}{\emph{RV parameters}} \\
$\ln{\lambda/\text{day}}$ & $4.75^{+0.22}_{-0.10}$ & - \\
$\ln{\Gamma}$ & $-0.04^{+1.9}_{-1.9}$ & - \\
$\ln{P_{\text{rot}}/\text{day}}$ & $3.82^{+0.10}_{-0.11}$ & - \\
$\ln{a_{\text{HARPS-N}}/\text{m/s}}$ & $2.94^{+0.71}_{-0.69}$ & - \\
$\ln{a_{\text{HIRES}}/\text{m/s}}$ & $1.46^{+0.76}_{-0.61}$ & - \\
Jitter, $s_{\text{HARPS-N}}$ [\mps{]} & $1.18^{+0.64}_{-0.75}$ & $1.37^{+0.46}_{-0.40}$ \\
Jitter, $s_{\text{HIRES}}$ [\mps{]} & $0.11^{+0.61}_{-0.09}$ & $2.47^{+1.10}_{-0.83}$ \\
Velocity offset, $\gamma_{\text{HARPS-N}}$ [\mps{]} & $-0.81^{+2.81}_{-3.03}$ & $1.39^{+0.45}_{-0.43}$ \\
Velocity offset, $\gamma_{\text{HIRES}}$ [\mps{]} & $0.69^{+2.50}_{-2.70}$ & $-0.33^{+0.96}_{-0.99}$ \\
\smallskip \\
\multicolumn{3}{c}{\emph{TOI-1235 b parameters}} \\
Orbital period, $P$ [days] & $3.444729^{+0.000031}_{-0.000028}$ &  $3.444727^{+0.000035}_{-0.000039}$ \\
Time of mid-transit, $T_0$ [BJD - 2,457,000] & $1845.51696^{+0.00099}_{-0.00098}$ & $1845.5173^{+0.0008}_{-0.0010}$ \\
Transit duration $D$ [hrs] & $1.84^{+0.09}_{-0.16}$ & $1.94^{+0.05}_{-0.04}$ \\
Transit depth, $Z$ [ppt] & $0.645^{+0.049}_{-0.044}$ & $0.662^{+0.039}_{-0.038}$ \\
Scaled semimajor axis, $a/R_s$ & $13.20^{+0.41}_{-0.40}$ & $13.15^{+0.34}_{-0.32}$ \\
Planet-to-star radius ratio, $r_p/R_s$ & $0.0254\pm 0.0009$ & $0.0257\pm 0.0007$ \\
Impact parameter, $b$ & $0.45^{+0.21}_{-0.19}$ & $0.33^{+0.15}_{-0.19}$ \\
Inclination, $i$ [deg] & $88.1^{+0.8}_{-0.9}$ & $88.6^{+0.8}_{-0.6}$ \\
$e\cos{\omega}$ & - & $0.00^{+0.03}_{-0.03}$ \\
$e\sin{\omega}$ & - & $0.00^{+0.04}_{-0.06}$ \\
$\sqrt{e}\cos{\omega}$ & $0.07^{+0.13}_{-0.15}$ & - \\
$\sqrt{e}\sin{\omega}$ & $-0.02^{+0.23}_{-0.23}$ & - \\
Eccentricity, $e$ & $<0.15$\tablenotemark{c} & $<0.16$\tablenotemark{c} \\
Planet radius, $r_p$ [\Rearth{]} & $1.738^{+0.087}_{-0.076}$ & $1.763^{+0.071}_{-0.066}$ \\
Log RV semi-amplitude, $\ln{K/m/s}$ & $1.41^{+0.10}_{-0.13}$ & $1.46^{+0.11}_{-0.13}$ \\
RV semi-amplitude, $K$ [\mps{]} & $4.11^{+0.43}_{-0.50}$ & $4.32^{+0.50}_{-0.51}$ \\
Planet mass, $m_p$ [\Mearth{]} & $6.91^{+0.75}_{-0.85}$ & $7.53^{+0.88}_{-0.89}$ \\
Bulk density, $\rho_p$ [g cm$^{-3}$] & $7.4^{+1.5}_{-1.3}$ & $7.5^{+1.4}_{-1.2}$ \\
Surface gravity, $g_p$ [m s$^{-2}$] & $22.6^{+3.5}_{-3.4}$ & $23.7^{+3.3}_{-3.2}$ \\
Escape velocity, $v_{\text{esc}}$ [km s$^{-1}$] & $22.4^{+1.3}_{-1.5}$ & $23.1^{+1.2}_{-1.1}$ \\
Semimajor axis, $a$ [AU] & $0.03845^{+0.00037}_{-0.00040}$ & $0.03846^{+0.00033}_{-0.00032}$ \\
Insolation, $F$ [F$_{\oplus}$] & $53.6^{+5.3}_{-4.7}$ & $53.6^{+4.2}_{-4.3}$ \\
Equilibrium temperature, $T_{\text{eq}}$ [K] && \\
\hspace{2pt} Bond albedo = 0.0 & $754\pm 18$ & $754\pm 18$ \\
\hspace{2pt} Bond albedo = 0.3 & $689\pm 16$ & $689\pm 16$ \\
\multicolumn{3}{c}{\emph{Keplerian parameters of the 22-day RV signal}\tablenotemark{d}} \\
Period [days] & - &  $21.99^{+0.47}_{-0.32}$ \\
Reference epoch (analogous to $T_0$) [BJD - 2,457,000] & - & $1835.34^{+0.89}_{-0.87}$ \\
Log RV semi-amplitude, $\ln{K/m/s}$ & - & $1.50^{+0.15}_{-0.14}$ \\
RV semi-amplitude, $K$ [\mps{]} & - & $4.50^{+0.62}_{-0.57}$ \\
$e\cos{\omega}$ & - & $0.02^{+0.11}_{-0.09}$ \\
$e\sin{\omega}$ & - & $0.09^{+0.18}_{-0.10}$ \\
\enddata
\tablenotetext{a}{Our fiducial model features sequential modeling of the \tess{} light curve followed by the RV analysis conditioned on the results of the transit analysis.}
\tablenotetext{b}{Our alternative analysis is a global model of the \tess{} and ground-based light curves, along with the RVs using the \texttt{EXOFASTv2} software.}
\tablenotetext{c}{95\% upper limit.}
\tablenotetext{d}{The 22-day RV signal is modeled as an eccentric Keplerian in our \texttt{EXOFASTv2} model although we emphasize that here we do not attribute this signal to a second planet.}
\end{deluxetable*}

\suppressAffiliationsfalse
\allauthors
\end{document}